\documentclass[12pt,preprint]{aastex}

\def\fenix{PHOENIX}
\def\cloudy{CLOUDY}
\def\wmb{\emph{WM-Basic}}
\def\te{$T_e$}
\def\tenii{$T_e$(N$^+$)}
\def\teoii{$T_e$(O$^+$)}
\def\teoiii{$T_e$(O$^{+2}$)}
\def\ne{$N_e$}
\def\nh{$n(H)$}
\def\nesii{$N_e$([S II])}
\def\cm{cm$^{-3}$}
\def\hii{H$^+$}
\def\nii{N$^+ $}

\def\oii{O$^+$}
\def\oiii{O$^{+2}$}
\def\sii{S$^+$}
\def\fnii{[N~II]}
\def\foii{[O~II]}
\def\foiii{[O~III]}
\def\fsii{[S~II]}
\def\oiih{12 + log(O/H)}
\def\niio{N$^+$/O$^+$}
\shorttitle{N/O IN LOW METALLICITY SYSTEMS}
\shortauthors{Nava et al.}
\begin{document}
\title{ON THE DETERMINATION OF N AND O ABUNDANCES IN LOW METALLICITY SYSTEMS}
\author{Aida Nava, Darrin Casebeer, Richard B. C. Henry, and Darko Jevremovic}
\affil{Homer L. Dodge Department of Physics and Astronomy, The University of Oklahoma, 440 West Brooks, Room 131, Norman OK, 73019-2061}
\email{nava,casebeer,henry,djc@nhn.ou.edu}
\begin{abstract}
We show that in order to minimize the uncertainties in the N and O abundances of low mass, low metallicity (O/H $\le$ solar/5) emission-line galaxies, it is necessary to employ separate parameterizations for inferring \tenii~and \teoii~from \teoiii. In addition, we show that for the above systems, the ionization correction factor (ICF) for obtaining N/O from \niio, where the latter is derived from optical emission-line flux ratios, is $\left< ICF \right>$ = 1.08 $\pm$ 0.09. These findings are based on state-of-the-art single-star H II region simulations, employing our own modeled stellar spectra as input. Our models offer the advantage of having matching stellar and nebular abundances. In addition, they have O/H as low as solar/50 (lower than any past work), as well as log(N/O) and log(C/O) fixed at characteristic values of -1.46 and -0.7, respectively. The above results were used to re-derive N and O abundances for a sample of 68 systems with \oiih~$\le$ 8.1, whose de-reddened emission-line strengths were collected from the literature. The analysis of the log(N/O) versus \oiih~diagram of the above systems shows the following: (1) the largest group of objects forms the well-known N/O plateau with a value for the mean (and its statistical error) of -1.43 (+.0084/-.0085); (2) the objects are distributed within a range in log(N/O) of -1.54 to -1.27 in Gaussian fashion around the mean with a standard deviation of $\sigma=$+.071/-.084; and (3) a $\chi$--square analysis suggests that only a small amount of the observed scatter in log(N/O) is intrinsic.
\end{abstract}
\section{INTRODUCTION}
Low metallicity systems (or metal-poor systems) are low mass emission-line galaxies with \oiih~roughly in the range 7.2-8.1, where O/H is the oxygen to hydrogen number density ratio. They are referred to in the literature as dwarf irregular galaxies (dIs), their bursting derivatives, blue compact galaxies (BCGs), or H II galaxies, a subclass of BCGs dominated in the optical by H II region-like spectra \citep[see][]{kunth00}. We use oxygen as a measure of metallicity in the above systems because it is the most abundant metal in the interstellar medium (ISM), and because its abundance can be easily determined from the emission-lines present in their optical spectra. Note that if we adopt \oiih$_\odot$ = 8.66~$\pm$ 0.05 for the solar value \citep{asplund04}, then low metallicity systems have O/H in the range solar/30-solar/4. The lower limit corresponds to the value of I Zw 18, which is the most metal-poor dwarf galaxy ever observed. The upper limit ensures that the most metal rich objects are still below the onset of the observed N/O increase with O/H. In addition, objects with \oiih~below 8.1 have sufficient strength in the auroral line \foiii~ $\lambda4363$, which is necessary in order to determine the electron temperatures required to obtain N/O.  

In the log(N/O) versus \oiih~diagram (e.g., Fig. 2 of \citealt{pilyugin04}), low metallicity systems form what is known as the primary N/O plateau. For these objects, N/O is independent of O/H as is expected for the ratio of two primary elements\footnote{An element is referred to as primary if its production process is independent of the initial metallicity of the progenitor star.}. The analysis of the properties of this plateau plays an important role in understanding the process of primary nitrogen production, which is essential for determining the stellar mass range most responsible for N-production, constraining stellar N-yields (in particular for massive stars), obtaining information about the initial mass function (IMF) of galaxies, and perfecting galactic chemical evolution models. Note that although one could think of other candidates for studying N/O at low metallicities, such as stars in metal-poor globular clusters, Galactic halo stars, or damped Ly$\alpha$ systems, which are quasar radiation absorbers with O/H $\sim$ solar/10, it is much harder to measure N and O abundances in such systems. This explains why the bulk of the N/O data are from low metallicity systems.

There is disagreement among authors on the vertical thickness of the plateau and on the origin of its properties. \cite{pagel85} called attention to the large scatter in N/O at fixed O/H. However, \cite{thuan95}, \cite{izotov99}, and \cite{izotov01}, found that the plateau is narrow, with a dispersion of only $\pm$ 0.02 dex in log(N/O), for objects with \oiih~$\leq$ 7.6. In addition, the latter authors explained the plateau in terms of primary nitrogen from rapidly evolving massive stars (M $>$ 8 M$_\odot$). In contrast, \cite{henry00} and \cite{chiappini03} concluded that these objects are dominated by primary nitrogen from intermediate mass stars, which have masses in the range 4-8 M$_\odot$ and evolve much slower. 

In order to correctly interpret the plateau properties, it is important to up-date the techniques employed to determine the N/O values. This can be accomplished through modeling. In particular, simulations are required to obtain N/O for two reasons.

Firstly, they are necessary for finding parameterizations for the electron temperatures of the regions where N$^+$ and O$^+$ are emitting, i.e., \tenii~and \teoii. Unfortunately, because the strengths of the auroral features \fnii~$\lambda5755$ and \foii~$\lambda\lambda7320,7230$ are commonly unavailable in observational data, the above temperatures cannot be obtained from emission-line flux ratios of the form, nebular line(s) / auroral line(s), although they are required in order to determine the abundances of the above two ions. Since photoionization models can provide the auroral lines missing from observations, these lines can be used in turn to compute \tenii, \teoii, and \teoiii, independent of each other. Because objects with oxygen abundances below 8.1 usually have sufficient strength in the auroral line [O III] $\lambda4363$, \teoiii~can be directly determined from the observed temperature sensitive line flux ratio $(I_{4959}+I_{5007})/I_{4363}$ \citep{shields81}. In general, a parameterization of the form \tenii~= \teoii~= \emph{f}[\teoiii], inferred from models, is used to compute N/O at low O/H (e.g., \citealt{kobulnicky96,izotov99,melbourne04}). Examples of this type of relation, based on photoionization models by \cite{stasinska90}, can be found in \cite{pagel92} or \cite{izotov94}. A more recent parameterization based on models by \cite{stasinska96} is used in \cite{stasinska03}, although the stellar spectra used as input to these models reflect the state-of-the-art in 1995. In addition, the lowest stellar atmosphere metallicity that they used is solar/10. Ever since, more complete synthetic stellar spectra have been published for metallicities as low as solar/20 (e.g., \citealt{smith02}). Parameterizations of the form \tenii~=~\emph{f$_1$}[\teoiii] and \teoii~=~\emph{f$_2$}[\teoiii] for systems with metallicities as low as that of I Zw 18 should be obtained and compared to each other, using results from up-to-date stellar and nebular models. 

Secondly, models are required for finding the ionization correction factor (ICF) necessary for obtaining N/O from N$^+$/O$^+$. Based on simulations described in Garnett \& Shields (1987), which make use of Mihalas (1972) and Kurucz (1975, 1979) stellar atmospheres with effective temperatures ranging from 38 kK to 55 kK, \cite{garnett90} showed that 0.8 $\lesssim$ (N/O)/(\niio) $\lesssim$ 1.0 in ionized nebulae with one-tenth \citep{anders89} solar abundances. More recently, \cite{izotov04} confirmed that N/O $\approx $ \niio~is a reasonable approximation in metal-poor H II galaxies. They established this using photoionization models of \cite{stasinska03} with input stellar spectra generated with the stellar atmosphere code \emph{CoStar} \citep{schaerer97}, as well as simulations computed with the evolutionary synthesis code Starburst99 \citep{leitherer99}, using \cite{smith02} stellar spectra generated with \wmb~\citep{pauldrach01}. Note that the spectra of \cite{smith02} are available for metallicities down to Z$_\odot$/20. Unfortunately, the uncertainty in the ICF is not given in \cite{izotov04}, however, the value of the ICF is sensitive to the ionization parameter\footnote{The ionization parameter can be defined as U $\equiv$ Q(H) / [ 4 $\pi$ r$_o^2$ n(H) c ], where, r$_o$ is the distance from the ionizing source to the illuminated face of the cloud, n(H) is the total hydrogen density, c is the speed of light, and Q(H) is the stellar emission rate of hydrogen ionizing photons \citep{ferland98}.}. This parameter is a poorly constrained variable (e.g., log(U) ranges from -5 to -1 in models by \citealt{garnett89}). Note that more recently, \cite{stasinska96} found that only a small range for U can fit the emission-line diagnostics from the few H II galaxies that they modeled. However, as these authors conclude, U is yet to be observationally constrained. In addition, the value of the ICF also depends on the method employed to compute \niio. In particular, if the latter is derived from optical line flux ratios, the ICF will depend on the temperatures used to derive the \nii/\hii~and \oii/\hii~ratios.

Our goal is to re-derive N and O abundances for a carefully selected sample of 68 low-metallicity systems whose de-reddened optical emission-line strengths were taken from the literature (\citealt{campbell86,walsh89,pagel92,skillman93,izotov94,thuan95,kobulnicky96,izotov97,kobulnicky98,izotov98a,izotov98b,izotov99,melbourne04}). This was accomplished using independent temperature parameterizations for \tenii~and \teoii~(which diverge at low metallicities, see below) and applying an ICF for obtaining N/O from \niio. Our temperature patameterizations and ICF are based on results from a representative grid of photoionization models computed with the code \cloudy~version 96.01 \citep{ferland98}. These employ state-of-the-art modeled stellar spectra that we generated with \fenix~version 13.08.04A \citep{hauschildt99, hauschildt04, aufdenberg02}. The purpose of our work is to carefully study the morphology of the N/O plateau, assess the uncertainites, and then quantify the amount of natural scatter among plateau objects. 

We discuss our general procedure for deriving the number density ratios O/H and N/O from optical emission-lines in \S~\ref{sec:gprocedure}. In \S~\ref{sec:conditions} we describe our models and explain how we obtained the electron temperatures, electron density, and ICF relevant to the above abundance determinations. Our method for determining abundance uncertainties is explained in \S~\ref{sec:abundances}, in which we also compare our abundances with literature results. In section \S~\ref{sec:analysis} we analyze the scatter in the N/O plateau. Finally, section~\ref{sec:conclusion} gives a summary of our work. In a companion paper to this one (Henry et al. 2006), chemical evolution models are combined with Monte Carlo techniques to further interpret the plateau morphology.
\section{METHOD}
\subsection{General Procedure}\label{sec:gprocedure}
The N and O ionic abundances for our sample of low metallicity systems were determined directly from published optical emission-lines using the general expression:
\begin{equation}
\frac{ X^i }{ H^+ }= \frac{ I_{\lambda} }{ I_{H\beta} }~\frac{ N_e~\alpha^{eff}_{H\beta}(T_e) }{ \chi^u(T_e, N_e)~A^u_l }~\frac{ \lambda }{ 4861 },\label{eq:ionabun}
\end{equation}
where the fraction on the left is the number density ratio of the N or O ion relative to H$^+$, the first fraction on the right is the line flux ratio of a nebular forbidden feature of ion X$^i$ relative to the strength of H$\beta$, \ne~is the electron density (cm$^{-3}$), \te~is the electron temperature of the region where the relevant ion is emitting (K), $\alpha^{eff}_{H\beta}$ is the effective recombination coefficient of H$\beta$ which includes radiative and three-body processes (cm$^3$ s$^{-1}$), $\chi^u$ is the fraction of ions X$^i$ with an electron in the upper level of the transition of interest, A$^u_l$ is the corresponding spontaneous de-excitation rate coefficient (s$^{-1}$), and the last term is the wavelength ratio of the line of interest and H$\beta$ ({\AA}). Note that equation~(\ref{eq:ionabun}) is based on the assumption that H$\beta$ arises from recombination. The contribution to H lines from collisional excitation was neglected because the excitation potentials of H levels are much higher than the average thermal equilibrium temperature that characterizes H II regions \citep{osterbrock06}. In addition, equation~(\ref{eq:ionabun}) assumes that nebular forbidden lines originate from collisionally excited levels. Collisional excitation is significant in this case because the low-lying energy levels of the relevant ions are of the order of \emph{k}T. However, according to \cite{rubin86}, recombinations of O$^{+2}$ can excite the nebular doublet \foii~$\lambda\lambda3726, 3729$, used to compute \oii/\hii. The effect of this process will be analyzed in the future.

Although the presence  of the nebular He II  4686 \AA~emission-line in the  spectra of several metal-poor systems implies the  presence of unobserved O$^{+3}$/H$^+$, the contribution of this ion to the total oxygen abundance amounts to only a few percent according to our photoionization models. Therefore, we obtained the oxygen abundance by assuming that O/H = ( O$^+$ + O$^{+2}$ ) / \hii. The N/O ratio was determined from N/O = ICF $\times$ N$^+$/O$^+$, where the ICF was derived using models as explained in \S~\ref{sec:conditions}.  

A 5$^{th}$ order fit to data from \cite{storey95} was adopted for $\alpha^{eff}_{H\beta}$, i.e., $\alpha^{eff}_{H{\beta}}$ = $10^{\xi}$, where $\xi~=~\sum_{i=0}^5\ a_i$\te$^i$. The parameters are: $a_{0}=-12.404592$, $a_{1}=-3.47193796\times10^{-4}$, $a_{2}=4.98365006\times10^{-8}$, $a_{3}=-3.77545451\times10^{-12}$, $a_{4}=1.33944026\times10^{-16}$, and $a_{5}=-1.75120267\times10^{-21}$. The fraction $\chi^u$ was determined using a 5-level atom routine written and tested thoroughly for accuracy by one of us (A. N.). The equilibrium equations for each electronic level were solved using Gauss-Jordan elimination and the maximum pivot technique (Chapman 1998). Column (1) of Table~\ref{atomicdata} shows the list of ions involved in the determination of N/O. Columns (2), (3), and (4) of the same table give references for the energy levels, transition rates, and collisional strengths, respectively, employed in the 5-level atom calculations. Note that \sii~is included in Table~\ref{atomicdata} since \nesii~was employed in our calculations as explained in the next section.
\subsection{Electron Temperatures, Density, and ICF}\label{sec:conditions}

\subsubsection{Electron Temperatures}

In equation~(\ref{eq:ionabun}), three parameters, $\alpha^{eff}_{H\beta}$, $~\chi^u$, and \ne, depend on \te. In theory, \tenii, \teoii, and \teoiii~can be obtained from the \te~sensitive line flux ratios, $R$\fnii$\equiv(I_{6548}+I_{6583})/I_{5755}$, $R$\foii$\equiv(I_{3726}+I_{3729})/(I_{7320}+I_{7230})$, and $R$\foiii$\equiv(I_{4959}+I_{5007})/I_{4363}$, respectively \citep{shields81}. Unfortunately, in practice \tenii~and \teoii~cannot be directly inferred from their corresponding ratios because the auroral line \fnii~$\lambda5755$ and the auroral doublet \foii~$\lambda\lambda7320,7230$ are usually absent in observational data. However, photoionization models can predict the auroral line strengths required for deriving theoretical relations of the form \tenii~=~$f_1$[\teoiii] and \teoii~=~$f_2$[\teoiii]. The independent variable of these relations is \teoiii, since this temperature can be inferred directly from the observed $R$\foiii~ratio of low metallicity systems as long as \foiii~$\lambda4363$ is detected with good S/N. Note that the advantage of finding individual parameterizations for \tenii~and \teoii~ is that this minimizes the uncertainty in \niio~(and N/O) which is introduced by assuming that \tenii~and \teoii~are equivalent functions of \teoiii, as discussed in \S~1.

In order to find separate temperature parameterizations for \tenii~and \teoii, determine the ICF for obtaining N/O from \nii/\oii, and carefully estimate uncertainties in N/O, we computed a grid of photoionization models, representative of low metallicity systems. In the following paragraphs, we introduce \fenix, the stellar atmosphere code used to generate the spectra that served as input to our \cloudy~photoionization models, we explain the choice of parameter space covered by our stellar and nebular simulations, we describe how we proceeded to obtain the temperature parameterizations and ICF from our models, and we derive the electron temperatures, electron densities, and ionic abundances of our sample objects.  

\fenix~is a general stellar atmosphere code which operates in one-dimensional spherically symmetric or plane-parallel geometry. It can handle static atmospheres (e.g. stars, Jupiter-like planets, brown dwarfs) as well as moving atmospheres (supernovae, novae, massive hot stars, etc.). The code simultaneously solves the equations of radiative transfer (in the proper geometry), structure, and level population in Non-Local Thermodynamic Equilibrium (NLTE). 

We used \fenix's wind module package \citep{aufdenberg02} to generate synthetic spectra in the range from 100 to 8$\times$10$^6$ {\AA}, for six windy O-dwarf stars with solar/5, solar/20, or solar/50 metal abundances, except for N and C, which we scaled in order to be consistent with typical proportions found in low metallicity systems, i.e., log(N/O) = -1.46 and log(C/O) = -0.7 \citep{henry00}. The set of solar abundances was modified in order to match that of the code we used for our photoionization models\footnote{Our software uses solar abundances by \cite{allende01}.}. Other input parameters to the models were based on those employed by \cite{smith02}, who generated \wmb~synthetic spectra for the same type of stars, with solar/5 and solar/20 proportions for all metals (including N and C). 

We adopted a $\beta$-law for the wind velocity, i.e., 
\begin{equation}
v(r)~=~v_{\infty}~(~1~-~R_*/r~)^{\beta}, 
\end{equation}
where r is the distance from the center of the star, $R_*$ is the radius of the ``photosphere'' (considered to be the base of the wind), and $v_{\infty}$ is the terminal velocity, assumed to be reached at 100 $\times$ $R_*$. The $\beta$-index was assumed to be 0.8 \citep{lamers99}. The density structure in the dynamic region ($v(r)>0$) was calculated from the continuity equation:
\begin{equation}
\rho(r)~=~\frac{\dot{M}}{4~\pi~r^2~v(r)},
\end{equation}
where, $\rho(r)$ is the mass density and $\dot{M}$ is the mass loss rate. In the hydrostatic region ($v(r)=0$) the density structure was calculated assuming hydrostatic equilibrium. All models have 45 zones for the dynamic region and five zones for the hydrostatic region. 

\begin{figure}
\epsscale{.80}
\plotone{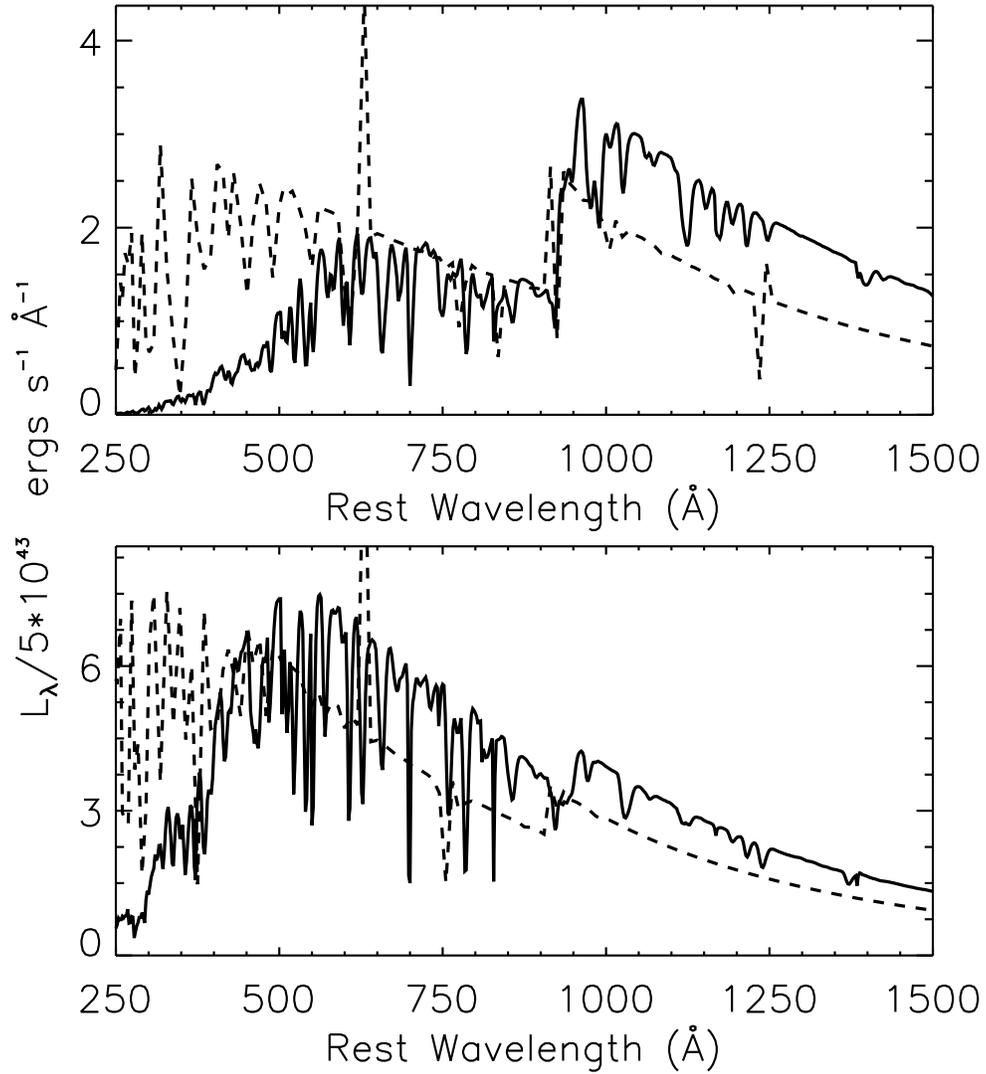}
\caption{Modeled spectra generated with \fenix~(solid lines) and with \wmb~(dashed lines) for two O-dwarfs having O/H = solar/20. The top panel features a 43 kK star, while the bottom panel features a 50 kK star.\label{fenix.vs.wmb}}
\end{figure}

A great deal of acceleration comes from resonance line radiation pressure of metals. Therefore, as in \cite{smith02}, the dependences of $v_\infty$ and $\dot{M}$ on the metallicity were parametrized by:
\begin{equation}
v_\infty~=~v_{\infty,\odot}~Z^{0.13},
\end{equation}
and
\begin{equation}
\dot{M}~=~\dot{M}_\odot~Z^{0.8}.
\end{equation}
Values for v$_{\infty,\odot}$ and $\dot{M}_\odot$ can be found in \cite{smith02}. A summary of the parameters characterizing our \fenix~models is given in Table~\ref{stars}. For each model listed in column (1), columns (2) through (8) show the following information: (2) atmosphere's metallicity, (3) stellar mass loss rate, (4) wind's terminal velocity, (5) photospheric temperature, (6) surface gravity, (7) stellar radius, and (8) total luminosity in the range specified within parenthesis.   
In Figure~\ref{fenix.vs.wmb}, we display the ionizing portion of \fenix~(solid lines) and \wmb~(dashed lines) spectra for two solar/20 O-dwarfs. The effective temperatures of the stars are 43 kK (top panel) and 50 kK (bottom panel). Note that the \fenix~models differ from the \wmb~models in that the former have more hydrogen ionizing photons and fewer helium ionizing photons. This is due to the fact that \fenix~absorbs the He ionizing flux due to the presence of more lines. Therefore, the normalized luminosity is re-distributed towards H ionizing energies.

We used \cloudy~to simulate the spectra of 24 single-star H II regions. The PHOENIX star used as input to each model is listed in column (1) of Table~\ref{clouds}. Note that we reduced the spectral resolution of the \fenix~energy distributions in order to fit \cloudy's default resolution. This was done using the CONGRID routine implemented in IDL version 5.5\footnote{This Solaris sunos sparc data analysis software was developed by Research Systems Inc. (2001).}. 

The following general characteristics are shared by all of our models: spherical geometry (ionizing source fully covered), constant H density \nh, homogeneous distribution of the gas (filling factor of 1), and static configuration. Studying the effects of using more complex models falls beyond the scope of this paper.

All nebulae have He/H = 0.1 and metal abundances identical to those of the atmospheres of the illuminating stars, except for the refractory elements Na, Mg, Al, Si, Ca and Fe, which are depleted by a factor of 10 since they are likely to be found in the form of grains in H II regions \citep{savage96}. Accordingly, grains were included in our simulations by turning on the ``grains ISM'' feature implemented in \cloudy, with their abundances scaled down by a factor of 1/5, 1/20, or 1/50, in order to be consistent with the oxygen depletion of the model\footnote{Note that \cloudy's ISM grains are based on an older set of solar abundances than the set used by the code to scale the H II region abundances. The latter set is based on the abundances given by \cite{grevesse98}, \cite{holweger01}, and \cite{allende01,allende02}}.

In order to study the effect of varying the nebula's electron density, we ran models for two different values of \nh, 10 or 100 \cm, which are typical values measured in low metallicity systems. We also varied the distance from the star to the illuminated face of the cloud (r$_0$), adopting either 2 pc, (which is bigger than the radius of a planetary nebula) or 5 pc (which allowed us to cover a range for the ionization parameter comparable to what is found in the literature). The above choices resulted in -3.17 $\leq$ log(U) $\leq$ -0.42. The parameters characterizing our \cloudy~models are summarized in Table~\ref{clouds}, which gives O/H in column (2), the H density in column (3), r$_0$ in column (4), log(U) in column (5) and the nebular geometry in column (6). Note that the latter is plane-parallel if $\Delta r/r_o < 0.1$, a thick shell if $0.1 < \Delta r/r_o < 3$, and spherical if $\Delta r/r_o\geq 3$, where $\Delta r$ is the thickness of the cloud set by the stopping temperature, chosen to be 100 K.

In order to find parameterizations of \tenii~and \teoii~with respect to \teoiii, from models, \tenii, \teoii, and \teoiii~need to be determined independent of each other first. This can be accomplished using two different methods. The first method, corresponding to common practice \citep{garnett90,stasinska90,stasinska96,stasinska03}, is to compute each temperature from: 
\begin{equation}
\left<T_e(A^{+j})\right>=\frac{ \int T_e(A^{+j}) N_e N(A^{+j})\mathrm{d}V }{ \int N_e N(A^{+j})\mathrm{d}V},\label{eq:meante}
\end{equation}
i.e., each temperature is an ion and \ne~weighted volume \te~mean. However, the above three temperatures can also be obtained from their respective line flux ratios, $R$[N II], $R$[O II], or $R$[O III], as explained below. This second method has the advantage of being consistent with the observational way of obtaining \teoiii, i.e., from line flux ratios. Ideally, temperature parameterizations should be independent of the method employed to compute the individual temperatures. We will compare temperatures computed with different methods further on, but first, we will explain how we determined \tenii, \teoii, and \teoiii~from their respective line flux ratios. 

In order to accomplish the latter, we used an iterative procedure based on Brent's method for finding roots \citep{press96}. For each of the 24 photoionization models, the idea is to use our 5-level atom routine to compute repeatedly the theoretical line flux ratio, $R_{the}$, for assumed values of \te~over a range of 4-25~kK until it matches the model-predicted value $R_{mod}$ to within a tolerance of 0.001. We carried out this procedure separately for the two density regimes of 10 and 100~cm$^{-3}$. Using a maximum of 100 iterations to solve the equation numerically, we searched for the value of \te~which satisfied the relation:
\begin{equation}
\frac{ R_{the}[T_e,N_e]~-~R_{mod} }{ R_{the}[T_e,N_e] }~=~0.001.\label{eq:zbrent}
\end {equation}

Henceforth, we will refer to models with \nh~= 10 \cm~and \nh~= 100 \cm~as the n10 and n100 models, respectively. Our temperature parameterizations for the n10 models are
\begin{equation}
T_{e}(N^+)=-5950+2.256\times T_{e}(O^{+2})-6.28\times10^{-5}\times T_{e}(O^{+2})^2\label{eq:tn210}
\end{equation}
and
\begin{equation}
T_e(O^+)=-5470+2.131\times T_{e}(O^{+2})-5.54\times10^{-5}\times T_{e}(O^{+2})^2,\label{eq:to210}
\end{equation}
while for the n100 models they are
\begin{equation}
T_{e}(N^+)=-7720+2.438\times T_{e}(O^{+2})-6.56\times10^{-5}\times T_{e}(O^{+2})^2,\label{eq:tn2100}
\end{equation}
and
\begin{equation}
T_e(O^+)=-7330+2.325\times T_{e}(O^{+2})-5.82\times10^{-5}\times T_{e}(O^{+2})^2.\label{eq:to2100}
\end{equation}

\begin{figure}
\epsscale{.80}
\plotone{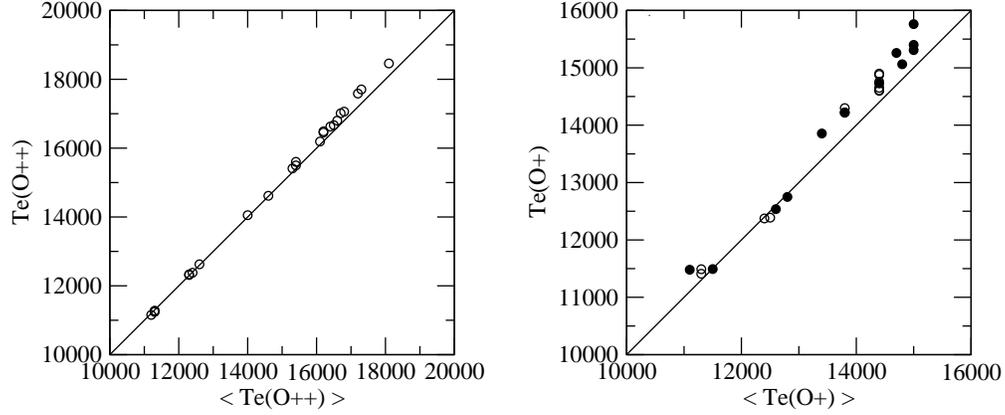}
\caption{Left: \teoiii~from $R$[O III] (vertically) vs. \teoiii~from eq. [\ref{eq:meante}] (horizontally) for each model. Right: Parameterized \teoii~(vertically) vs. \teoii~from eq. [\ref{eq:meante}] (horizontally). For the parameterized \teoii, we used eq. [\ref{eq:to210}] for the n10 models (open circles) and eq. [\ref{eq:to2100}] for the n100 models (closed circles). \label{meante-ratiote}}
\end{figure}

For each model, Figure~\ref{meante-ratiote} shows two independent values of \teoiii~on left panel, and two independent values of \teoii~on the right panel. On the left panel, the value obtained using $R$[O III] is given vertically, while the value $\left<T_e(O^{+2})\right>$ predicted by equation (\ref{eq:meante}) is given horizontally. The agreement between the two \teoiii~temperatures is excellent. On the right panel, the parameterized \teoii~given by either equation (\ref{eq:to210}) or (\ref{eq:to2100}) [depending on the input value of n(H)] is given vertically, with open circles representing n10 models and filled circles representing n100 models. On the same panel, the horizontal value is that predicted by equation (\ref{eq:meante}). The agreement between the two \teoii~temperatures is reasonable, with the parameterized value being higher than $\left<T_e(O^+)\right>$ for most models. The biggest difference found is \teoii~- $\left<T_e(O^+)\right>$ of about 800 K. The effect of this difference on the \niio~ratios is discussed below. Since unfortunately \cloudy~does not print out $\left<T_e(N^+)\right>$, the latter temperature was not compared to the parameterized \tenii.

\begin{figure}
\epsscale{.80}
\plotone{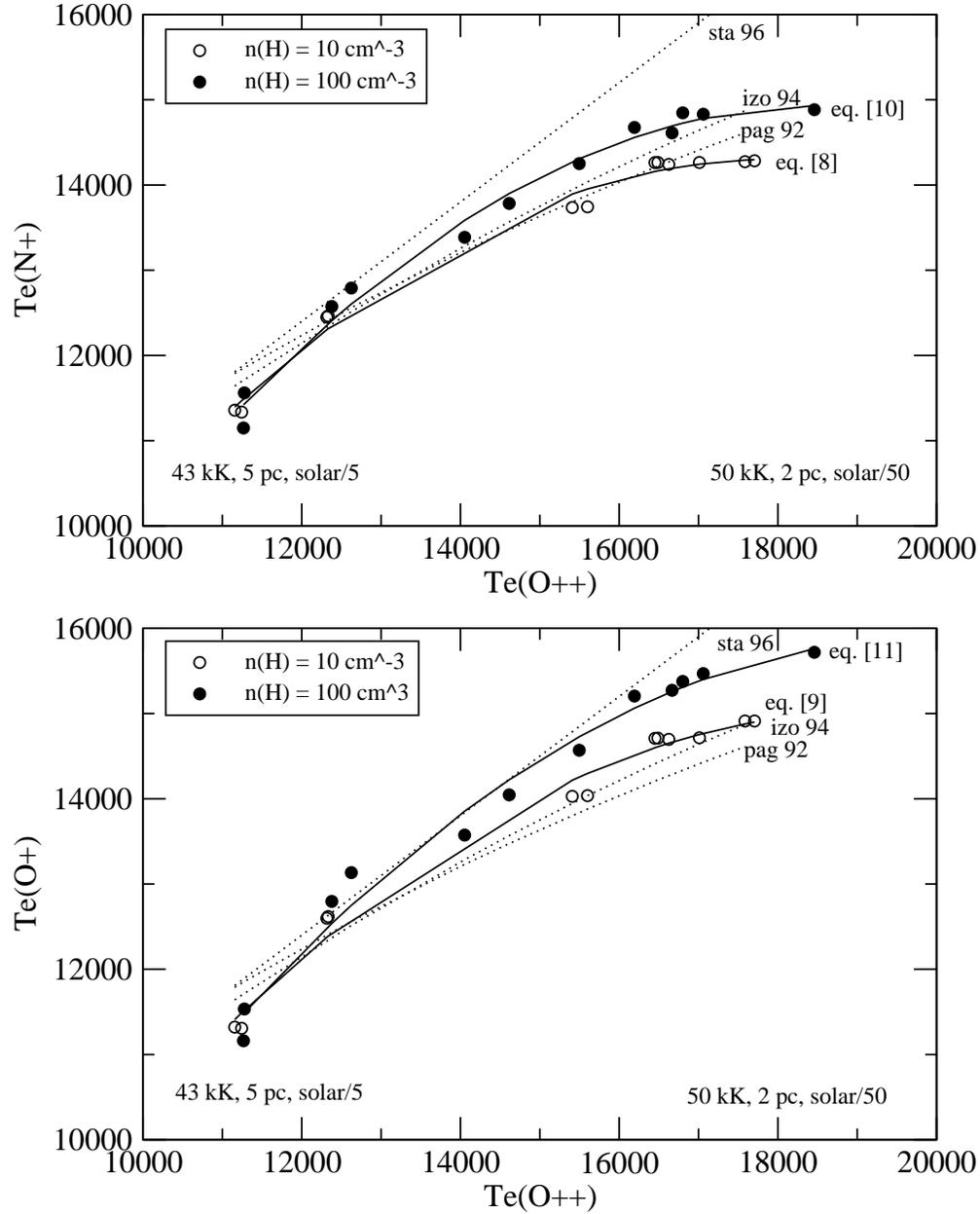}
\caption{Electron temperature parameterizations given by our models (solid lines), and by pag92, izo94, and sta96 (dotted lines). The solid lines labeled eq. [\ref{eq:tn210}] (top panel) and eq. [\ref{eq:to210}] (bottom panel) are best fits to n10 models (open circles). The solid lines labeled eq. [\ref{eq:tn2100}] (top panel) and eq. [\ref{eq:to2100}] (bottom panel) are best fits to n100 models (closed circles). $T_{eff}$, $r_o$, and O/H are indicated for the models located at the two \teoiii~extrema.\label{teplot}}
\end{figure}

Figure~\ref{teplot} features \tenii~vs. \teoiii~(top panel) and \teoii~vs. \teoiii~(bottom panel), where results for the n10 and n100 models are shown with open and filled circles, respectively. In this figure, the solid lines are best fits to models sharing the same value of \nh~while the dotted lines are extrapolations to low metallicities of parameterizations of the form \tenii~= \teoii~= $f$[\teoiii], from \cite{pagel92}, \cite{izotov94}, and \cite{stasinska96}. Note that the latter three fits are based on temperatures given by equation (\ref{eq:meante}). Henceforth we will refer to them as pag92, izo94, and sta96.

\begin{figure}
\epsscale{.80}
\plotone{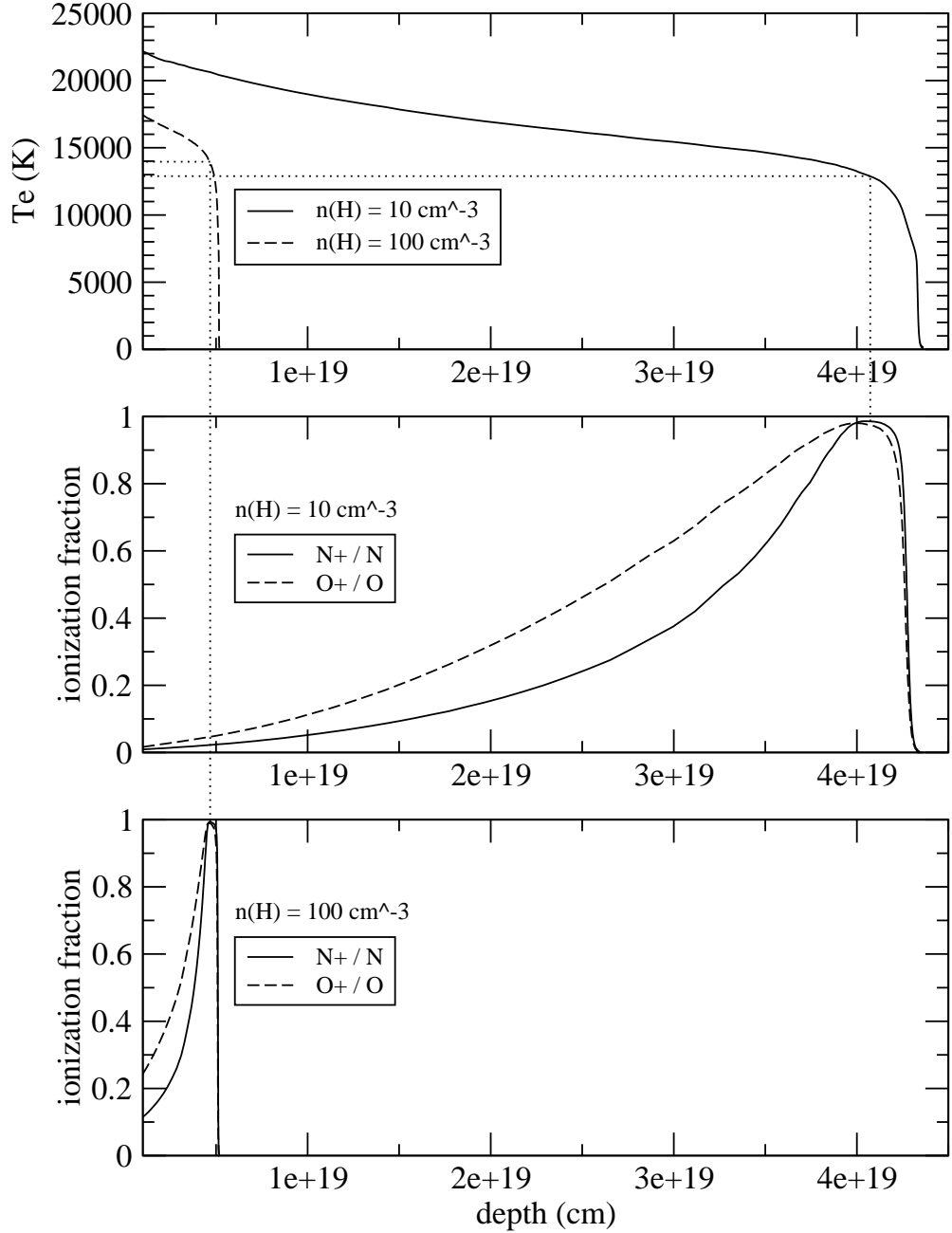}
\caption{Example of how \te, \nii/N, and \oii/O vary with depth as n(H) increases. The two models shown have $T_{eff}$ = 43 kK, $r_o$ = 2 pc, O/H = solar/50, and they differ only in their value of \nh. The solid and dashed lines in the top panel give the \te-structure for the n10 and n100 models (respectively), while they correspond to \nii/N and \oii/O in the middle (n10 model) and bottom (n100 model) panels. The dotted lines lead to \te~at $Max$\{\nii/N\} for each model.\label{tevsne}}
\end{figure}

The following can be noticed when analyzing Figure~\ref{teplot}. For a given ion (\nii~or \oii), \te~(vertical axis) at a given \teoiii~is in general higher for the n100 models than for the n10 models. This is due to the fact that as \nh~increases, U decreases, bringing the regions where \nii~and \oii~are emitting closer to the ionizing source, making them hotter. This behavior is illustrated in Figure~\ref{tevsne}, which has three panels. The top panel shows \te~as a function of depth from the illuminated face of the cloud for two models with $T_{eff}$ = 43 kK, $r_o$ = 2 pc, and O/H = solar/50, only differing in their value of \nh. The lower two panels show the ionization fractions \nii/N (solid line) and \oii/O (dashed line) as functions of depth for the two models. The middle panel corresponds to the n10 model, while the bottom panel corresponds to the n100 model. The dotted lines clearly show that \te~at the peak of \nii/N is higher by roughly 1000~K for the n100 model. The same is true for \te~at the peak of \oii/O. 

\begin{figure}
\epsscale{.80}
\plotone{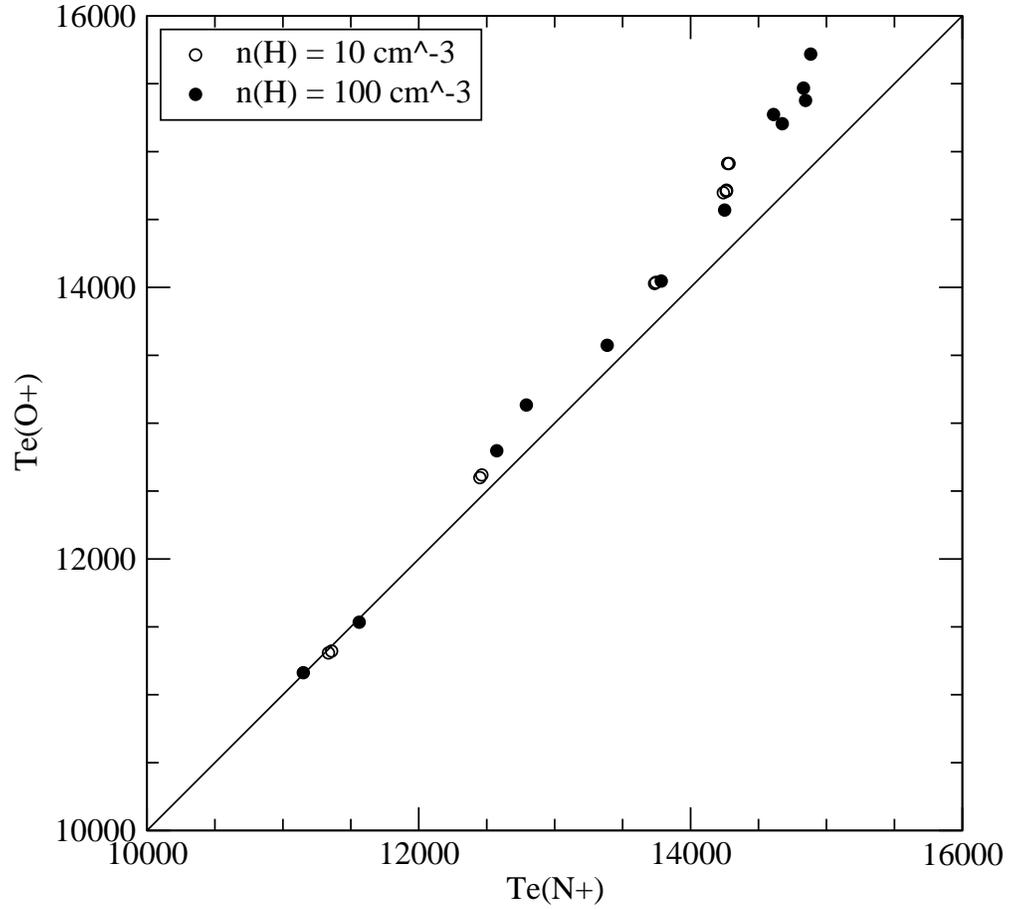}
\caption{Diagram showing \teoii~vs. \tenii, where the temperatures were obtainted using line flux ratios from our 24 models. Models with \nh~= 10 \cm~are shown with open circles, while models with \nh~= 100 \cm~are shown with filled circles. The diagonal corresponds to points such that \tenii~= \teoii.\label{to2vstn2}}\end{figure}

Also in Figure~\ref{teplot} we see that for a given value of \nh, \tenii~is in general lower than \teoii~[except for the two models with $T_{eff}$ = 43 kK, $r_o$ = 2 pc, and O/H = solar/5, for which they are almost identical]. This is shown more clearly in Figure~\ref{to2vstn2}, which features \tenii~vs. \teoii~for all models, and is a result of the fact that the regions where \nii~and \oii~emit do not completely overlap, the \oii~region extending more towards the star into hotter gas, as we pointed out earlier in our discussion of Figure~\ref{tevsne}. Our models also indicate that this offset becomes larger with rising stellar temperature and falling metallicity.

\begin{figure}
\epsscale{.60}
\plotone{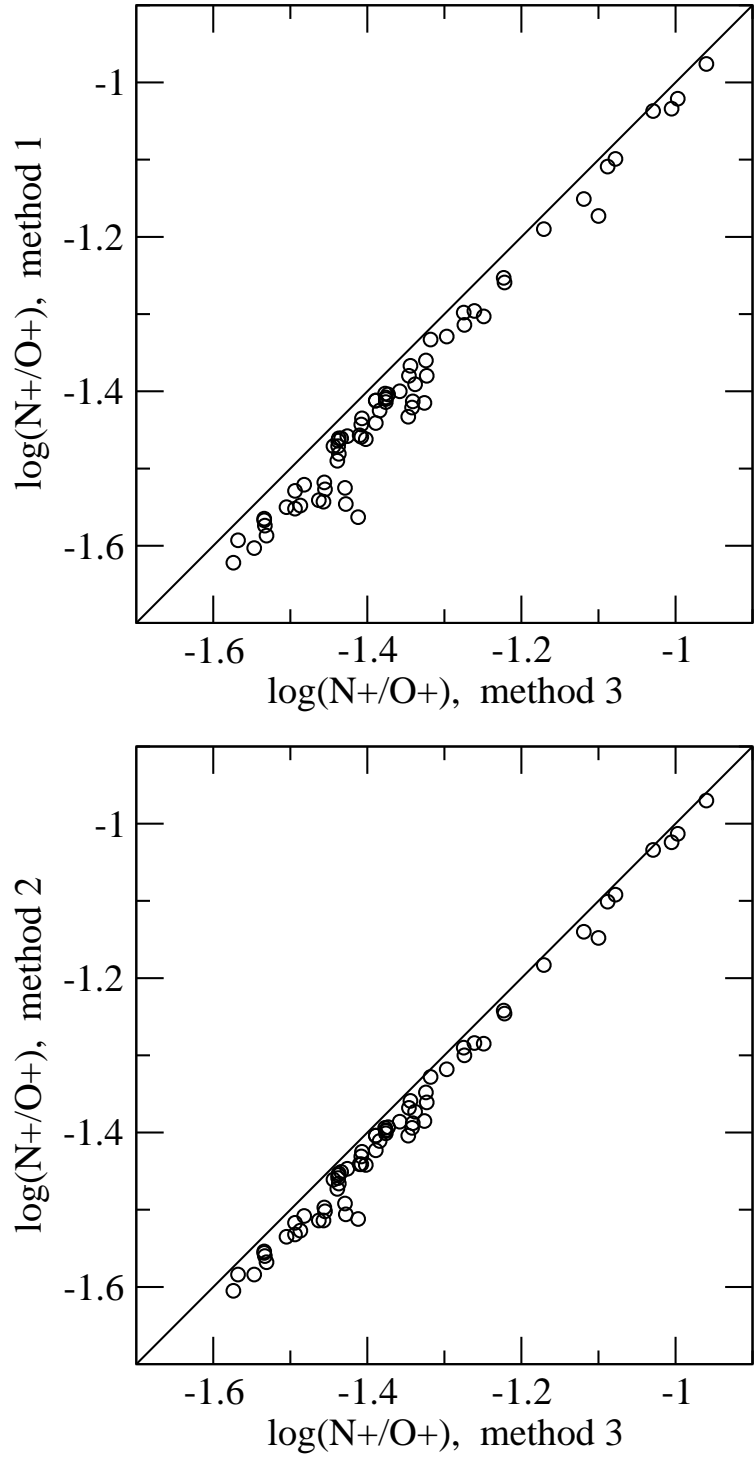}
\caption{Comparison of \nii/\oii~ratios derived for our sample objects. The top panel shows method 1 vs. method 3, while the bottom panel shows method 2 vs. method 3.\label{tn2andorto2}}
\end{figure}

In order to show that it is better to use \tenii~independent of \teoii~when determining the \niio~ratios of low metallicity systems, we derived the \niio~ratios of our sample objects using three different methods but always adopting \ne~= 100 \cm. Methods 1 and 2 both assume that \tenii~= \teoii, the assumption frequently made by researchers in the past. In method~1, both temperatures are computed using equation (\ref{eq:tn2100}), while in method~2 equation (\ref{eq:to2100}) is used instead. Then values of \nii/\hii, \oii/\hii, and ultimately \niio are determined from these derived temperatures. Finally, method 3 uses equations (\ref{eq:tn2100}) and (\ref{eq:to2100}) to compute \nii/\hii, \oii/\hii, respectively. The resulting three sets of \niio~ratios are shown in Figure~\ref{tn2andorto2}. The top panel is a plot of method 1 versus method 3, while the bottom panel is a plot of method 2 versus method 3. We see that using either method 1 or 2 results in systematically underestimating the \niio~ratios by roughly .05-.10~dex. Therefore, in order to minimize uncertainties in determining \niio~in low metallicity systems, one clearly should derive \tenii~and \teoii~from \teoiii~separately.
 
\begin{figure}
\epsscale{.80}
\plotone{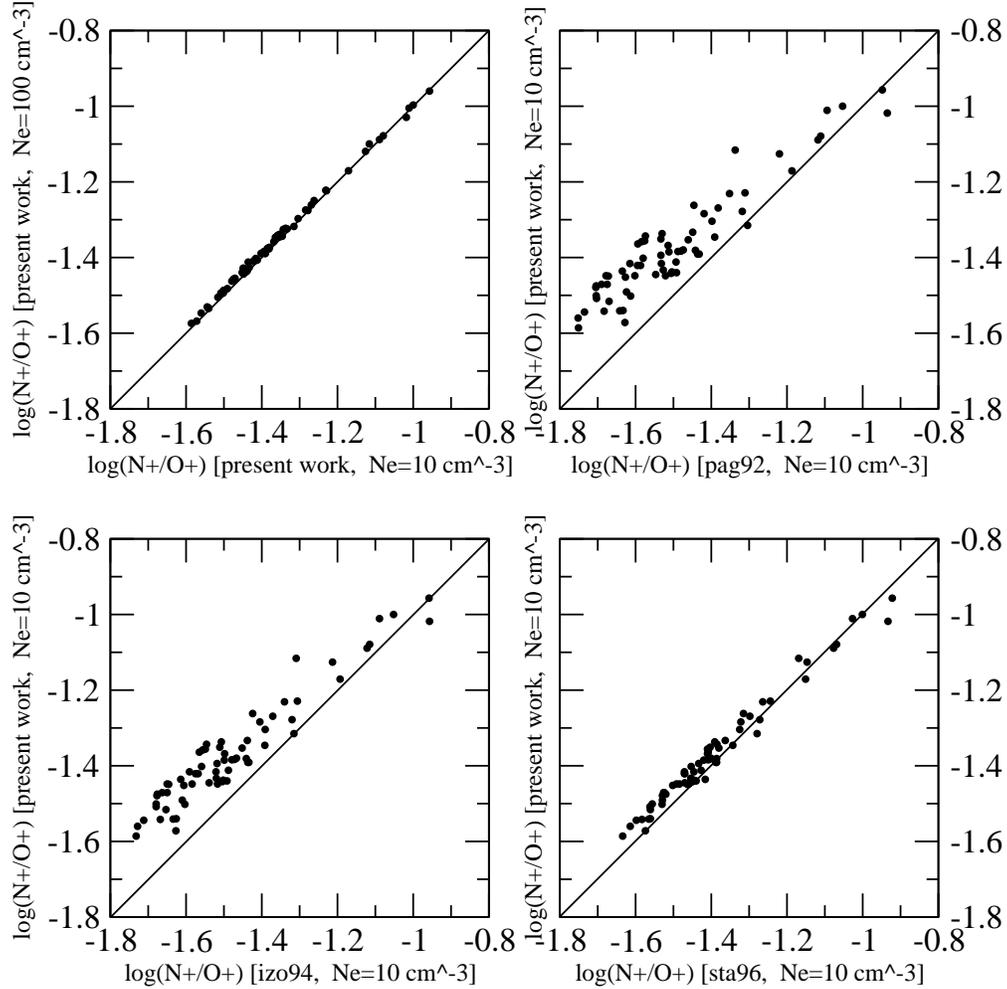}
\caption{Comparison of log(\niio)~values derived for our sample objects using various parameterizations. The top-left panel is a plot of values obtained using Eqs. [\ref{eq:tn2100}] \& [\ref{eq:to2100}] (vertical axis) vs. values obtained using Eqs. [\ref{eq:tn210}] \& [\ref{eq:to210}] (horizontal axis). In the remaining panels values obtained using Eqs. [\ref{eq:tn210}] \& [\ref{eq:to210}] (vertical axis) are plotted against values obtained using pag92, izo94, or sta96, adopting \ne~= 10 \cm.\label{n2os}}
\end{figure}

A comparison of our temperature parameterizations (eqs. [\ref{eq:tn210}] and [\ref{eq:to210}]) with those of pag92 and izo94 in Figure~\ref{teplot} shows good agreement among the three functions at a density of 10 \cm, where the pag92 and izo94 formulations are based on photoionization models by \cite{stasinska90}, most of which have \nh~= 10 \cm. On the other hand, for a density of 100~cm$^{-3}$ the agreement between our results (eqs. [\ref{eq:tn2100}] and [\ref{eq:to2100}]) and pag92 and izo94 is less satisfactory. For the temperature formulation by sta96, there is relatively poor agreement with our low density results for \tenii~and \teoii, although the agreement is better for our results for a density of 100~cm$^{-3}$, especially in the case of \teoii. Differences between the \niio~values obtained with sta96 and our method are due to the fact that sta96 constitutes an extrapolation to lower metallicites of the published relation. In addition, the latter is based on simulations of H II regions created by an evolving group of stars, while we use single non-evolving star simulations. Finally, our method uses independent relations for \tenii~and \teoii with respect to \teoiii.
To test the impact of the density effect on our final \niio~values, we used equation (\ref{eq:ionabun}) to calculate this ratio for each of our sample objects for the two density regimes, n10 and n100. The results are shown in the upper left panel of Figure~\ref{n2os}, where the n100 regime (ordinate) is plotted against the n10 regime (abscissa).
Clearly, the points closely follow the diagonal line corresponding to equal values of \niio~in both cases. 

Further comparisons of our temperature method with the three literature results just discussed are shown in 
the remaining panels of Figure~\ref{n2os}. In each case we have calculated \niio~values at a density of 10~cm${-3}$ for our sample objects using both our temperature method (vertical axis) and that of pag92 (upper right), izo94 (lower left), and sta96 (lower right), shown on the horizontal axis in each graph. The diagonal lines show the one-to-one relation. Note the particularly poor agreement between our method and those of pag92 and izo94, especially for \niio~values below -1.2. This disagreement can be easily traced to the relatively low \teoii~temperatures predicted by pag92 and izo94 when \teoiii~exceeds 14000~K, resulting in higher \oii/H$^+$ and lower \niio~ratios than ours. On the other hand, the apparently closer agreement with sta96 arises because their parameterizations for both \tenii~and \teoii~are offset from our curves by roughly the same amount in each case, resulting in relatively little difference in our computed \niio~ratios.

Since in general the only \ne~sensitive line flux ratio available from observations is the \fsii~ratio $I_{6716}$/$I_{6731}$, we assumed that \ne~was equal to \nesii~throughout the observed H II regions. For 42 low metallicity systems in our sample the \fsii~ratio was detected with reasonable S/N. For this group of objects, \nesii~was computed by applying Brent's method as described above, but replacing $R_{mod}$ with the observed value, $R_{obs}$. We proceeded as follows. We first fixed \ne~at 100 \cm~(a typical value for H II regions) and used $R_{obs}$\foiii~and equation (\ref{eq:zbrent}) to determine \teoiii. Next, from \teoiii,  $R_{obs}$\fsii~and equation (\ref{eq:zbrent}) we computed \ne. This new value of \ne~was then employed in the derivation of the N and O abundances of the above 42 objects. For the remaining objects, we used the average \ne~found for this subsample, i.e., $\sim$200 \cm.

Our final values for \ne, \teoiii, \teoii, and \tenii~are listed for each object in our sample in columns (2)  through (5) in Table~\ref{ionabun}.

Finally, we used our models to derive an ICF for obtaining N/O from \niio. This was accomplished by computing the average of (N/O)/(\niio) for the n100 models, where relevant line fluxes and parameterizations were employed to obtain each \niio~ratio, while N/O was fixed and given by the input value to our models, i.e., log(N/O) = -1.46. The average ICF for these models is 1.08 with a standard deviation of 0.09. This contrasts with common practice which is to use ICF = 1.

Unfortunately, the ICF depends on the method employed to compute \niio. A common practice is to compute \niio~from the ion fractions $\left<N^+/N\right>$ and $\left<O^+/O\right>$ given by:
\begin{equation}
\left< \frac{A^{+j}}{A_{tot}} \right> \equiv \frac{\int A^{+j} N_e \mathrm{d}V}{\int A_{tot} N_e \mathrm{d}V},\label{eq:meanifrac}
\end{equation}
and the input values of N and O. A comparison between the \niio~ratios that we used to compute our final ICF, and those given by equation (\ref{eq:meanifrac}), is shown in Figure~\ref{icf}. In this figure, the difference between log(\niio) and log($\left< N^+/O^+ \right>$) is plotted against log($\left< N^+/O^+ \right>$). The latter value is that predicted by equation (\ref{eq:meanifrac}), while log(\niio) is the value obtained from relevant line flux ratios and the temperature parameterizations for the n100 models. Note that Figure~\ref{icf} illustrates the need of an ICF for obtaining N/O at low metallicities, no matter which method is used to compute it. Clearly, log($\left< N^+/O^+ \right>$) deviates from log(N/O) which as mentioned above is fixed at -1.46 for all models. In fact, the use of $\left< N^+/O^+ \right>$ when computing the ICF results in an average of 1.28 with a standard deviation of 0.1. 

The vertical difference in Figure~\ref{icf} is at most $\sim$0.11 dex. It is positive because \niio~values from line flux ratios and parameterized temperatures are higher than those predicted by equation (\ref{eq:meanifrac}). This fact is reflected in the smaller ICF obtained when using line flux ratio \niio~values. Since our method is consistent with the way in which \niio~is measured observationally, i.e., from line flux ratios, we applied an ICF of 1.08 when deriving the N/O abundances of our sample objects. The ICF's contribution to the final N/O uncertainties was accordingly taken to be $\pm$0.09.

\section{IONIC AND ELEMENTAL ABUNDANCES}\label{sec:abundances}

We used the above results for density, temperatures, and ICF determinations to derive new values for the O/H and N/O ratios for our sample objects. The corresponding uncertainties were obtained using standard error propagation, where the quantities considered in our estimation of the abundance errors is given in Table~\ref{error}. The quantities listed in column (1) were assumed to be functions of the variables given in column (2). Note that because \ne~depends on \te~and visa versa, and since in addition, \ne~had to be assumed for a significant number of objects (see below), \ne~does not appear in Table~\ref{error}.

For each object in our sample, Table~\ref{ionabun} gives the object name in column (1), \ne~in column (2), \teoiii, \teoii, and \tenii~in columns (3), (4), and (5), respectively, and O$^+$/H$^+$, O$^{+2}$/H$^+$, and N$^{+}$/H$^+$ in columns (6), (7), and (8), respectively. Our final abundance results are given in Table~\ref{bigtable}, which lists object name in column (1), references for the observed emission-line strengths in column (2), our derived values for \oiih~$\pm~\sigma$ and log(N/O) $\pm~\sigma$ in columns (3) and (4), and literature values and references in columns (5), (6), and (7), respectively. 

Figures~\ref{oldnew_o2h} and~\ref{oldnew_n2o} feature comparisons of our O/H and N/O values (vertical axes) against published values. The diagonals show points of one-to-one correspondence. Our O/H values are systematically lower with respect to past calculations due 
primarily to our new temperature scheme for calculating O$^+$/H$^+$. However, in general, the error bars show agreement between our results and literature values. Our N/O values are also offset (higher by $\sim$ 0.05 to 0.1 dex) due the use of new temperature parametrizations and an ICF for obtaining N/O.

In Figure~\ref{oldnew_sigma_n2o}, we plot our uncertainties in log(N/O) against corresponding literature values. Some of the published uncertainties were estimated using Monte Carlo simulations to propagate the errors in the relevant line strengths \citep{campbell86, kobulnicky96}, and thus points occupy both sides of the diagonal due to the random nature of the latter technique. We point out that log(N/O) values with small uncertainties generally correspond to objects with small uncertainties in relevant emission-line strengths. The tendency for our error estimates to systematically exceed published ones is primarily due to the contribution of ICF uncertainty for each of our objects. However, we strongly argue that inclusion of this additional factor produces more realistic uncertainties than those previously published.

\section{ANALYSIS}\label{sec:analysis}

Figure~\ref{oursample} is a plot of N/O versus O/H for the abundances and uncertainties listed in Table~\ref{bigtable}. Recall that
N/O is a ratio which measures differential nucleosynthesis, while O/H is a
metallicity gauge and measures the extent of chemical evolution in the
area immediately surrounding the observed emission-line region. Furthermore, each data point in 
Figure~\ref{oursample} provides a
snapshot of current conditions within a galaxy, each of
which arrived at its present position by following a path determined by
its own unique star formation history.

Figure~\ref{n2ohisto} shows the distribution of N/O values for our
sample, using a bin size of 0.02~dex. Two distinct groups of objects are
apparent. The larger population comprises objects extending over the entire O/H range considered here and possessing N/O values which appear to be distributed in near-Gaussian fashion between limits of -1.54 $\le$ N/O $\le$-1.27. These limits are indicated with dashed lines in Figures~\ref{oursample} and \ref{n2ohisto}.
The remaining objects with N/O $>$ -1.27 appear to belong to a smaller population which is concentrated at the high end of our O/H range and possess relatively high N/O values compared with the larger group.

The 52 objects making up the larger population form what we will define
for purposes of analysis as the {\it N/O plateau}. The N/O plateau has been
discussed by numerous authors, including \cite{garnett90, thuan95},
\cite{kobulnicky96}, and \cite{pilyugin03}. We argue that by treating this group of objects separately, we are not doing so in an arbitrarily fashion. Rather we are led to this result for the following reasons. First, the distribution of the entire sample in the N/O dimension shown in Figure~\ref{n2ohisto} strongly hints at the existence of two distinct groups of objects. Second, our current theoretical understanding of N production strongly suggests that we should expect a narrow, horizontal track at low metallcities where the production is primary. Both analytical and numerical models such as those presented in \cite{henry00} predict the presence of this track. Finally, a theoretical analysis of the N/O plateau shape by Henry, Nava, \& Prochaska (2006) using numerical chemical evolution models finds that an additional N production process triggered at O/H$>$7.8 is needed to explain the high N/O objects. Thus, we conclude that current observational and theoretical evidence suggests that what we have defined as the N/O plateau consists of objects which have been subjected to similar N production mechanisms, and that we are justified in excluding, for the time being, those objects in the smaller group which possess higher N/O ratios.

We now discuss some of the statistical aspects of our measurements. In doing so we note that all
means, errors, and standard deviations relating to our sample and given
below were initially computed in linear space and then converted to their
corresponding logarithmic quantities. This is not necessarily true for
values quoted from other papers, however.

The log of the weighted mean of N/O for the 52 objects lying on the plateau is
-1.43 (+.0084/-.0085), where the values in parenthesis represent the error
in the mean, not the standard deviation, and were derived from
observational uncertainties quoted in column (4) of
Table~\ref{bigtable}\footnote{Note that the error in the mean depends upon the uncertainties in the individual measurements as
well as the square root of the number of objects in the sample, being inversely related to
the latter (see \citealt{bevington03}).}. The value for plateau objects
with O/H below 7.8 is -1.44 (+.011/-.012), while above 7.8 it is measured
to be -1.41 (+.012/-.013), so there is no evidence of metallicity
dependence of the mean. These results are in good agreement with those of
\cite{izotov99}, who obtained -1.47 ($\sigma = \pm$.14) for their total
sample, -1.60 ($\sigma = \pm$.02) for their low metallicity (O/H $<$ 7.6)
objects, and -1.46 ($\pm$.14) for their high metallicity (O/H $>$ 7.6)
objects. Likewise, our results are in close agreement with
\cite{garnett90}, who found log(N/O) = -1.46 (+.10/-.13) for plateau objects.
Finally, the standard deviation about the weighted mean for our plateau objects,
expressed logarithmically, is +.071/-.084, nearly an order of magnitude
larger than the error in the mean.

An interesting question pertaining to the distribution of points in the
N/O-O/H plane concerns the nature of the observed N/O scatter (as measured
by the standard deviation) of the plateau objects and to
what degree it is related to errors in abundance measurements. That is to
say, is the spread in N/O associated with the plateau due mostly to
intrinsic or statistical scatter?
To investigate this point further, we assumed a Gaussian distribution in
the parent N/O plateau population, as suggested by the distribution shape in
Figure~\ref{n2ohisto}, and performed a $\chi$-square analysis of the
observed point distribution of log(N/O) for the plateau using our
calculated weighted mean value for log(N/O) of -1.43 along with the established
errors in the abundances, as given in column (4) of Table~\ref{bigtable}. We
found that the value for the reduced $\chi$-square\footnote{The reduced
$\chi$-square, $\chi^2_{\nu} = \frac{1}{N-1}\sum_i
\left\{\frac{1}{\sigma^2_i}[y_i - \mu]^2\right\}$, where $y_i$ and
$\sigma_i$ are the N/O value and uncertainty of the {\it i}th object, $\mu$ is
the sample average, and N is the number of sample objects. Within the
summation, the denominator represents the measurement errors.} for the 52
plateau objects is 1.244, consistent with a probability of about 90\% that
our sample of objects could not have been drawn randomly from a parent
population characterized by our calculated mean log(N/O), error, and standard deviation. In
other words, some additional scatter is needed to explain the observed spread
in N/O\footnote{Concerned that a few points with small observational
uncertainties could be artificially inflating the total $\chi$-square
value, we calculated each object's contribution to the total $\chi$-square
and found no points to be excessively high.}.

In order to estimate the magnitude of this additional scatter, we recalculated
reduced $\chi$-square values for the 52 plateau objects numerous times, each time adding
intrinsic scatter to the measurement errors in quadrature to simulate real
scatter among the objects, with a target value for the reduced
$\chi$-square of unity. Such a value was reached when $\sigma$ for the
intrinsic component was 0.024, or about 1/3 of the magnitude of the standard deviation and
the typical statistical uncertainties in column (4) of Table~\ref{bigtable}.
Therefore, the implication of this $\chi$-square analysis is that only a small portion of the scatter observed for the plateau objects is intrinsic. Considering that the many of the uncertainties associated with the line strengths taken from the literature for our analysis are only estimated and not determined in a rigorous fashion, we cannot conclude with any confidence that a significant portion of the vertical scatter in N/O is real. Rather, the issue can only be decided by using better measurements including well-determined uncertainties in the abundances.

\section{CONCLUSIONS}\label{sec:conclusion}

This paper has been devoted to a detailed review of the method for deriving N and O abundances in low mass emission-line galaxies with 12 + log(O/H)$\leq$ 8.1. These low metallicity systems are classified in the literature as dwarf irregular galaxies, blue compact galaxies, or H II galaxies. The principal motivation for our work was to estimate the contribution of the N/O uncertainties to the scatter among N/O plateau objects in the log(N/O) versus \oiih~diagram. To carry out this study, we selected a sample of 68 objects from the literature and used their published de-reddened emission-line strengths to calculate a homogeneous set of abundances, paying special attention to the determination of [N~II] and [O~II] electron temperatures for which the necessary but weak auroral lines are usually absent from the data. In such cases these temperatures, which are needed to compute the abundances of their associated ions (N$^+$ and O$^+$), must be inferred from parametric relations in terms of the more accessible [O III] temperature. We established theoretical relations for the relevant temperatures based on single-star H II region simulations (modeled with \cloudy) and employing our own input stellar spectra (modeled with \fenix). These relations were derived from line flux ratio temperatures in contrast with common practice, which is to use ion and \ne-weighted volume temperature means. We also used our models to infer an ICF for obtaining N/O from \niio, and to carefully estimate N/O uncertainites by applying standard error propagation methods. This ICF was derived from line flux ratio ionic abundances, also in contrast with common practice, which is to use \ne-weighted volume ion fraction means. Our method has the advantage of being consistent with the procedure employed to derive \teoiii~and \niio~from observations, i.e., from line flux ratios. Note that \niio~values from line flux ratios and parameterized temperatures are higher than those predicted by equation (\ref{eq:meanifrac}). However, this fact is taken care of by our value for the ICF.

We conclude the following from our studies:

\begin{enumerate}

\item \tenii~and \teoii~diverge significantly from each other in low metallicity regimes. Therefore, the assumption of previous authors that these two temperatures are equivalent appears to be invalid at low metallicities.

\item A realistic value for the \nii/\oii~ionization correction factor is 1.08$\pm$.09.

\item Our final N/O values are systematically larger than previous results, while we find that O/H is generally below other measurements. This outcome is due to our new temperatures and ICF determinations.

\item The majority (52) of objects in our survey form a relatively flat, horizontal system in the N/O-O/H plane, consistent with the idea that both N and O production are primary at low metallicity. These galaxies comprise what we define as the N/O plateau.

\item Plateau objects appear to form a Gaussian distribution in log(N/O) between -1.54 and -1.27 with the logarithm of the weighted arithmetic average, log(N/O), being -1.43 (+.0084/-.0085) and a standard deviation of +.071/-.084.

\item A $\chi^2$ analysis of the plateau objects indicates that only a small fraction of the observed scatter in N/O is intrinsic, although this conclusion remains tentative until line strength uncertainties, which are the largest contributors to the abundance uncertainties, can be rigorously determined for a large sample of objects.

\item The remaining 16 sample objects have log(N/O) values exceeding -1.27 and therefore reside above the N/O plateau as we define it here.

Abundances of N and O are expected to improve in accuracy and precision as better temperature determinations become possible through the use of large telescopes capable of obtaining high S/N spectral observations of the auroral lines of \nii~and \oii. Until that time, photoionization models of increased sophistication can go a long way toward reducing the uncertainty, as we have shown here.

\end{enumerate}

\acknowledgments

We are greatful to E. Baron, J. X. Prochaska, Schaerer D., T. Hoffmann, and G. J. Ferland for helpful discussions and comments. This research is supported by NSF grant AST 03-07118 and NASA grant NAG5-3505 to the University of Oklahoma.

\clearpage
\begin{deluxetable}{llll}
\tabletypesize{\scriptsize}
\setlength{\tabcolsep}{0.03in}
\tablecolumns{4}
\tablewidth{0in}
\tablenum{1}
\tablecaption{ATOMIC DATA}
\tablehead{
\colhead{Ion} &
\colhead{Energy Levels} &
\colhead{Transition Rates} &
\colhead{Collision Strengths (at 10$^4$ K)}}
\startdata
\nii & Nist & Nist & \citealt{lennon94}\\
\oii & Nist & Nist & \citealt{mcLaughlin94}\\
\oiii & Nist & Nist & \citealt{lennon94}\\
\sii & \hfill & \citealt{mendoza83} & \citealt{ramsbottom96}\\\hline
\enddata\label{atomicdata}
\end{deluxetable}
\clearpage
\begin{deluxetable}{ccccccccc}
\tabletypesize{\scriptsize}
\setlength{\tabcolsep}{0.07in}
\tablecolumns{9}
\tablewidth{0in}
\tablenum{2}
\tablecaption{PHOENIX MODELS}
\tablehead
{
\colhead{ Star } &
\colhead{ Metal }  &
\colhead{ log($\dot{M}$) } &
\colhead{ $v_\infty$ } &
\colhead{ $T_{eff}$ } &
\colhead{ log(g) } &
\colhead{ R/R$_\odot$ } &
\colhead{ log(L) }
\\
\hfill &  abundances\tablenotemark{a}  &  (M$_\odot$ yr$^{-1}$) & (km s$^{-1}$)& (10$^4$ K) & (cm s$^{-1}$) & (R$_\odot$=6.95$\times10^{10}$ cm) & (erg s$^{-1}$)}
\startdata
1 & solar/5 & -6.80 & 2330& 4.265 & 4.00 & 10.48 & 38.954 (-3.517/0.960)\\
2 & solar/20 & -7.28 & 1940& 4.260 & 4.00 & 10.48 & 39.080 (-3.517/0.960)\\
3 & solar/50 & -7.40 & 1727& 4.260 & 4.00 & 10.48 & 39.042 (-3.517/0.960)\\
4 & solar/5 & -6.41 &  2550& 5.000 & 4.00 & 9.80 & 39.560 (-2.040/0.960)\\
5 & solar/20 & -6.89 &  2130& 5.000 & 4.00 & 9.80 & 39.500 (-3.040/0.960)\\
6 & solar/50 & -7.21 & 1894 & 5.000 & 4.00 & 9.80 & 39.300 (-2.040/0.960) \\ 
\enddata
\tablenotetext{a}{Except those of N and C which in order to match typical values measured in low metallicity systems, were set such that log(N/O)=-1.46 and log(C/O)=-0.7.}\label{stars}
\end{deluxetable}
\clearpage
\begin{deluxetable}{cccccc}
\tabletypesize{\scriptsize}              
\setlength{\tabcolsep}{0.07in}
\tablecolumns{6}  
\tablewidth{0in}  
\tablenum{3}
\tablecaption{CLOUDY MODELS}  
\tablehead{
\colhead{ Star\tablenotemark{a}} &
\colhead{ \oiih } &
\colhead{ $n(H)$ } &       
\colhead{ $r_{0}$ } & 
\colhead{ log(U) } &
\colhead{Geometry}\\
\hfill & \hfill & (cm$^{-3}$) &  (pc) & \hfill & \hfill }   
\startdata  
1 & 8.0 & 10 & 2 & -1.37 & spherical\\
1 & 8.0 & 10 & 5 & -2.17 & thick shell\\
1 & 8.0 & 100 & 2 & -2.37 & thick shell\\
1 & 8.0 & 100 & 5 & -3.17 & plane-parallel\\
 
2 & 7.4 & 10 & 2 & -1.14 & spherical\\
2 & 7.4 & 10 & 5 & -1.94 & thick shell\\
2 & 7.4 & 100 & 2 & -2.14 & thick shell \\
2 & 7.4 & 100 & 5 & -2.94 & thick shell\\

3 & 7.0 & 10 & 2 & -1.16 & spherical\\
3 & 7.0 & 10 & 5 & -1.96 & thick shell\\
3 & 7.0 & 100 & 2 & -2.16 & thick shell \\
3 & 7.0 & 100 & 5 & -2.96 & plane-parallel\\

4 & 8.0 & 10 & 2 & -0.39 & spherical\\
4 & 8.0 & 10 & 5 & -1.19 & spherical\\
4 & 8.0 & 100 & 2 & -1.39 & thick shell\\
4 & 8.0 & 100 & 5 & -2.19 & thick shell\\

5 & 7.4 & 10 & 2 & -0.42 & spherical\\
5 & 7.4 & 10 & 5 & -1.22 & spherical\\
5 & 7.4 & 100 & 2 & -1.42 & thick shell\\
5 & 7.4 & 100 & 5 & -2.22 & thick shell\\

6 & 7.0 & 10 & 2 & -0.68 & spherical \\
6 & 7.0 & 10 & 5 & -1.48 & spherical \\
6 & 7.0 & 100 & 2 & -1.68 & thick shell\\
6 & 7.0 & 100 & 5 & -2.48 & thick shell\\
\enddata
\tablenotetext{a}{The nebular and stellar abundances are identical for all elements except for Na, Mg, Al, Si, Ca and Fe, whose abundances are $0.1\times$their stellar value.}
\label{clouds}
\end{deluxetable}
\clearpage
\begin{deluxetable}{ll}
\tabletypesize{\scriptsize}
\setlength{\tabcolsep}{0.07in}
\tablecolumns{2}
\tablewidth{0in}
\tablenum{4}
\tablecaption{ERROR PROPAGATION}
\tablehead
{
\colhead{ Quantity } &
\colhead{ Contributing Factors }}
\startdata
$R$\foiii &  $I_{4363}$, $I_{4959}$, $I_{5007}$ \\
\teoiii & $R$\foiii \\
\teoii & \teoiii \\
\tenii & \teoiii \\
\oiii/\hii & $I_{5007}$, \teoiii \\
\oii/\hii & $I_{3727}$, \teoiii \\
\nii/\hii & $I_{6584}$, \teoiii \\
O/H & \oii, \oiii \\
N/O & ICF, \nii/\hii, \oii/\hii \\
\enddata
\label{error}
\end{deluxetable}
\clearpage
\begin{deluxetable}{lllllccc}
\tabletypesize{\scriptsize}
\setlength{\tabcolsep}{0.07in}
\renewcommand{\arraystretch}{1.5}
\rotate
\tablecolumns{8}
\tablewidth{0in}
\tablenum{5}
\tablecaption{PHYSICAL CONDITIONS AND IONIC ABUNDANCES}
\tablehead{
\colhead{Object Names} &
\colhead{\ne} &
\colhead{\teoiii} & 
\colhead{\teoii} &
\colhead{\tenii} &
\colhead{O$^{+}$/H$^+$} &
\colhead{O$^{+2}$/H$^+$} &
\colhead{N$^{+}$/H$^+$}\\
\colhead{\hfill} &
\colhead{(cm$^{-3}$)} &
\colhead{(K)} &
\colhead{(K)} &
\colhead{(K)} &
\colhead{($\times$10$^5$)} &
\colhead{($\times$10$^5$)} &
\colhead{($\times$10$^5$)}}
\startdata
I ZW 18 NW, MRK 116, 0930+554               & $<$50 &$19000\pm{800  }$&$15800\pm{100   }$&$14900\pm{100   }$&$ 0.178\pm{0.008}$&$ 1.338\pm{0.080}$&$ 0.006\pm{0.000}$\\
I ZW 18 SE, MRK 116, 0930+554               & 200 &$16700\pm{1400}$&$15300\pm{500  }$&$14700\pm{300  }$&$ 0.365\pm{0.045}$&$ 1.539\pm{0.298}$&$ 0.011\pm{0.001}$\\
SBS 0335-052                                & 250 &$20800\pm{400  }$&$15900\pm{100   }$&$14600\pm{100  }$&$ 0.201\pm{0.003}$&$ 1.802\pm{0.021}$&$ 0.007\pm{0.002}$\\
SBS 0940+544 N                              & 150 &$22100\pm{600  }$&$15600\pm{100  }$&$14100\pm{300  }$&$ 0.364\pm{0.017}$&$ 1.971\pm{0.028}$&$ 0.014\pm{0.002}$\\
SBS 1159+545                                & 100 &$19700\pm{600  }$&$15900\pm{100   }$&$14900\pm{100   }$&$ 0.460\pm{0.014}$&$ 2.138\pm{0.076}$&$ 0.017\pm{0.002}$\\
SBS 1415+437                                & 50  &$18600\pm{300  }$&$15800\pm{100   }$&$14900\pm{100    }$&$ 0.725\pm{0.012}$&$ 1.996\pm{0.047}$&$ 0.025\pm{0.001}$\\
KISSR 85                                    & 750 &$18100\pm{1100}$&$15700\pm{200  }$&$14900\pm{100   }$&$ 0.819\pm{0.063}$&$ 2.258\pm{0.259}$&$ 0.033\pm{0.005}$\\
UGC 4483, 0832+699                          & 50  &$16700\pm{300  }$&$15300\pm{100   }$&$14700\pm{100   }$&$ 0.803\pm{0.023}$&$ 2.321\pm{0.086}$&$ 0.023\pm{0.001}$\\
TOL 1214-277                                & 200 &$19100\pm{700  }$&$15800\pm{100   }$&$14900\pm{100   }$&$ 0.158\pm{0.012}$&$ 3.265\pm{0.178}$&$ 0.007\pm{0.001}$\\
KISSR 1490                                  & 200 &$19300\pm{1100}$&$15900\pm{100   }$&$14900\pm{100   }$&$ 1.046\pm{0.049}$&$ 2.465\pm{0.210}$&$ 0.048\pm{0.008}$\\
SBS 1211+540                                & 150 &$17200\pm{200  }$&$15400\pm{100   }$&$14800\pm{100   }$&$ 0.485\pm{0.013}$&$ 3.605\pm{0.113}$&$ 0.016\pm{0.001}$\\
KISSR 1013, KISSB 211                       & 400 &$18200\pm{1100}$&$15700\pm{200  }$&$14900\pm{100   }$&$ 1.573\pm{0.117}$&$ 2.709\pm{0.306}$&$ 0.117\pm{0.010}$\\
SBS 1249+493                                & 200 &$16900\pm{700  }$&$15300\pm{200  }$&$14700\pm{200  }$&$ 0.827\pm{0.055}$&$ 3.517\pm{0.333}$&$ 0.026\pm{0.003}$\\
116+583 B                                   & 500 &$16800\pm{900  }$&$15300\pm{300  }$&$14700\pm{200  }$&$ 0.490\pm{0.040}$&$ 3.922\pm{0.491}$&$ 0.021\pm{0.006}$\\
VII Zw 403, 1124+792                        & $<$50  &$15600\pm{200  }$&$14800\pm{100  }$&$14300\pm{100   }$&$ 1.128\pm{0.030}$&$ 3.397\pm{0.130}$&$ 0.042\pm{0.002}$\\
TOL 1304-353                                & 200 &$18600\pm{700  }$&$15800\pm{100  }$&$14900\pm{100    }$&$ 0.258\pm{0.009}$&$ 4.527\pm{0.373}$&$ 0.011\pm{0.003}$\\
SBS 1420+544, CG 413                        & 200 &$18000\pm{400  }$&$15700\pm{100   }$&$14900\pm{100   }$&$ 0.402\pm{0.013}$&$ 4.419\pm{0.174}$&$ 0.014\pm{0.002}$\\
SBS 1205+557                                & 200 &$16200\pm{700  }$&$15100\pm{300  }$&$14600\pm{200  }$&$ 1.759\pm{0.121}$&$ 3.312\pm{0.345}$&$ 0.067\pm{0.005}$\\
SBS 1128+573                                & 200 &$17000\pm{600  }$&$15400\pm{200  }$&$14800\pm{100  }$&$ 0.668\pm{0.035}$&$ 4.587\pm{0.387}$&$ 0.026\pm{0.005}$\\
C 1543+091                                  & $<$50  &$16500\pm{500  }$&$15200\pm{200  }$&$14600\pm{100  }$&$ 0.448\pm{0.024}$&$ 4.812\pm{0.403}$&$ 0.026\pm{0.006}$\\
UM 461, SCHG 1148-020                       & 50  &$17200\pm{400  }$&$15500\pm{100  }$&$14800\pm{100   }$&$ 0.353\pm{0.016}$&$ 4.934\pm{0.290}$&$ 0.012\pm{0.001}$\\
I ZW 36, MRK 209, HARO 29,  1223+487        & 50  &$16300\pm{100   }$&$15100\pm{100   }$&$14600\pm{100   }$&$ 0.566\pm{0.004}$&$ 4.865\pm{0.048}$&$ 0.023\pm{0.001}$\\
SBS 1331+493 N                              & 150 &$16200\pm{300  }$&$15100\pm{100  }$&$14600\pm{100   }$&$ 0.723\pm{0.025}$&$ 4.862\pm{0.205}$&$ 0.026\pm{0.002}$\\
MRK 1434, SBS 1030+583                      & 200 &$15600\pm{100  }$&$14800\pm{100   }$&$14300\pm{100   }$&$ 0.849\pm{0.016}$&$ 4.929\pm{0.121}$&$ 0.026\pm{0.001}$\\
MRK 193, SBS 1152+579                       & 150 &$16400\pm{200  }$&$15200\pm{100   }$&$14600\pm{100   }$&$ 0.709\pm{0.016}$&$ 5.204\pm{0.145}$&$ 0.032\pm{0.002}$\\
MRK 71, NGC 2343, 0723+692 B                & 200 &$15000\pm{300  }$&$14600\pm{200  }$&$14100\pm{200  }$&$ 1.483\pm{0.071}$&$ 4.637\pm{0.285}$&$ 0.048\pm{0.003}$\\
MRK 600, 0248+042                           & 50  &$15900\pm{200  }$&$14900\pm{100   }$&$14500\pm{100   }$&$ 1.069\pm{0.025}$&$ 5.116\pm{0.168}$&$ 0.029\pm{0.001}$\\
MRK 71, NGC 2343, SBS 0749+568 B            & 200 &$15300\pm{800  }$&$14600\pm{400  }$&$14200\pm{300  }$&$ 1.507\pm{0.162}$&$ 4.980\pm{0.711}$&$ 0.065\pm{0.007}$\\
MRK 36, 1102+294                            & 550 &$14900\pm{400  }$&$14400\pm{200  }$&$14000\pm{200  }$&$ 0.808\pm{0.058}$&$ 5.722\pm{0.414}$&$ 0.031\pm{0.007}$\\
POX 4: C1148-203                            & $<$50  &$15100\pm{400  }$&$14500\pm{200  }$&$14100\pm{200  }$&$ 0.755\pm{0.054}$&$ 5.800\pm{0.425}$&$ 0.041\pm{0.002}$\\
MRK 1416, SBC 0917+527                      & 200 &$15200\pm{300  }$&$14600\pm{200  }$&$14200\pm{100  }$&$ 1.734\pm{0.072}$&$ 4.885\pm{0.268}$&$ 0.050\pm{0.002}$\\
MRK 71, NGC 2343, 0723+692 A                & 100  &$15900\pm{100   }$&$14900\pm{100   }$&$14500\pm{100   }$&$ 0.505\pm{0.003}$&$ 6.132\pm{0.054}$&$ 0.020\pm{0.001}$\\
SBS 1533+574 A                              & $<$50  &$14500\pm{600  }$&$14100\pm{400  }$&$13800\pm{300  }$&$ 2.431\pm{0.226}$&$ 4.573\pm{0.505}$&$ 0.112\pm{0.007}$\\
MRK 1486, SBS 1358+576                      & $<$50  &$14800\pm{200  }$&$14300\pm{200  }$&$14000\pm{100  }$&$ 1.575\pm{0.054}$&$ 5.530\pm{0.235}$&$ 0.096\pm{0.003}$\\
SBS 1331+493 A                              & 200 &$13600\pm{900  }$&$13500\pm{700  }$&$13300\pm{600  }$&$ 2.837\pm{0.506}$&$ 4.278\pm{0.803}$&$ 0.102\pm{0.013}$\\
KISSR 675, KISSB 187                        & 1000 &$15200\pm{1100}$&$14600\pm{600  }$&$14200\pm{500  }$&$ 1.356\pm{0.218}$&$ 5.901\pm{1.211}$&$ 0.048\pm{0.013}$\\
FAIRALL 30                                  & 100 &$14600\pm{300  }$&$14200\pm{200  }$&$13900\pm{100  }$&$ 1.079\pm{0.044}$&$ 6.418\pm{0.373}$&$ 0.059\pm{0.003}$\\
C 0840+120                                  & 200 &$14200\pm{400  }$&$13900\pm{300  }$&$13700\pm{200  }$&$ 1.540\pm{0.111}$&$ 6.080\pm{0.482}$&$ 0.064\pm{0.009}$\\
SBS 0926+606                                & 150 &$14400\pm{200  }$&$14100\pm{200  }$&$13800\pm{100  }$&$ 1.851\pm{0.072}$&$ 5.812\pm{0.262}$&$ 0.076\pm{0.002}$\\
I ZW 49, MRK 59, NGC 4861, ARP 209, 1256+351& 100 &$14600\pm{700  }$&$14200\pm{500  }$&$13900\pm{400  }$&$ 2.280\pm{0.278}$&$ 5.460\pm{0.767}$&$ 0.108\pm{0.010}$\\
KISSR 1194                                  & 50  &$14600\pm{400  }$&$14200\pm{200  }$&$13900\pm{200  }$&$ 1.976\pm{0.175}$&$ 5.935\pm{0.487}$&$ 0.078\pm{0.008}$\\
KISSR 396, KISSB 145, WAS 81                & 200 &$14100\pm{500  }$&$13900\pm{300  }$&$13600\pm{300  }$&$ 2.705\pm{0.232}$&$ 5.352\pm{0.514}$&$ 0.078\pm{0.005}$\\
UM 420, 0218+003                            & 200 &$13900\pm{800  }$&$13800\pm{600  }$&$13500\pm{500  }$&$ 2.754\pm{0.425}$&$ 5.384\pm{0.915}$&$ 0.267\pm{0.026}$\\
1437+370                                    & 200 &$14200\pm{300  }$&$13900\pm{200  }$&$13600\pm{200  }$&$ 1.511\pm{0.079}$&$ 6.699\pm{0.382}$&$ 0.056\pm{0.003}$\\
KISSR 1778                                  & 50  &$13100\pm{900  }$&$13100\pm{700  }$&$13000\pm{700  }$&$ 3.622\pm{0.816}$&$ 4.605\pm{0.968}$&$ 0.149\pm{0.027}$\\
UM 462, SCHG 1150-021                       & 200 &$14300\pm{200  }$&$14000\pm{100  }$&$13700\pm{100   }$&$ 1.527\pm{0.050}$&$ 6.841\pm{0.275}$&$ 0.055\pm{0.007}$\\
SBS 1222+614                                & 200 &$14300\pm{100  }$&$14000\pm{100   }$&$13700\pm{100   }$&$ 1.231\pm{0.026}$&$ 7.369\pm{0.181}$&$ 0.035\pm{0.001}$\\
TOL 1304-386                                & 100  &$14200\pm{300  }$&$13900\pm{200  }$&$13600\pm{200  }$&$ 1.073\pm{0.061}$&$ 7.781\pm{0.502}$&$ 0.082\pm{0.006}$\\
1533+469                                    & $<$50 &$13900\pm{500  }$&$13800\pm{400  }$&$13500\pm{300  }$&$ 2.456\pm{0.257}$&$ 6.514\pm{0.740}$&$ 0.149\pm{0.012}$\\
SBS 0907+543                                & 100 &$14500\pm{400  }$&$14200\pm{300  }$&$13900\pm{200  }$&$ 0.939\pm{0.069}$&$ 8.113\pm{0.689}$&$ 0.030\pm{0.005}$\\
UM 469                                      & 200 &$12900\pm{1200}$&$13000\pm{1000}$&$12800\pm{900  }$&$ 2.458\pm{0.710}$&$ 6.662\pm{1.858}$&$ 0.201\pm{0.039}$\\
1054+365, CG 798                            & 200 &$13900\pm{200  }$&$13700\pm{100  }$&$13500\pm{100  }$&$ 1.210\pm{0.044}$&$ 8.010\pm{0.315}$&$ 0.047\pm{0.002}$\\
MRK 1450, SBS 1135+581                      & 200 &$13200\pm{200  }$&$13200\pm{100  }$&$13000\pm{100  }$&$ 1.764\pm{0.063}$&$ 7.459\pm{0.258}$&$ 0.071\pm{0.002}$\\
MRK 1271, TOL 1053+064                      & 50  &$14200\pm{200  }$&$13900\pm{100  }$&$13600\pm{100  }$&$ 1.677\pm{0.055}$&$ 7.752\pm{0.285}$&$ 0.086\pm{0.003}$\\
NGC 5253-6                                  & 100  &$12700\pm{600  }$&$12800\pm{500  }$&$12600\pm{500  }$&$ 4.204\pm{0.662}$&$ 5.374\pm{0.793}$&$ 0.284\pm{0.029}$\\
MRK 1304, UM 448, SCHG 1139+006             & 100 &$12200\pm{600  }$&$12400\pm{500  }$&$12300\pm{500  }$&$ 4.559\pm{0.766}$&$ 5.090\pm{0.756}$&$ 0.496\pm{0.050}$\\
MRK 22, SBS 0946+558                        & 50  &$13500\pm{200  }$&$13500\pm{200  }$&$13300\pm{100  }$&$ 1.753\pm{0.078}$&$ 7.896\pm{0.354}$&$ 0.064\pm{0.003}$\\
1441+294, CG 1258                           & 200 &$13200\pm{800  }$&$13200\pm{600  }$&$13000\pm{500  }$&$ 1.969\pm{0.340}$&$ 7.709\pm{1.326}$&$ 0.087\pm{0.012}$\\
KISS 49, KISSB 94, CG 177                   & 100  &$13100\pm{800  }$&$13100\pm{700  }$&$13000\pm{600  }$&$ 4.490\pm{0.888}$&$ 5.229\pm{1.010}$&$ 0.239\pm{0.037}$\\
MRK 1409, SBS 0741+535                      & 550 &$13700\pm{700  }$&$13600\pm{500  }$&$13400\pm{400  }$&$ 3.673\pm{0.486}$&$ 6.132\pm{0.869}$&$ 0.121\pm{0.011}$\\
0948+532                                    & 100  &$13500\pm{300  }$&$13400\pm{200  }$&$13200\pm{200  }$&$ 1.631\pm{0.096}$&$ 8.250\pm{0.492}$&$ 0.069\pm{0.004}$\\
TOL 1345-420                                & 100  &$13400\pm{700  }$&$13400\pm{500  }$&$13200\pm{400  }$&$ 2.034\pm{0.279}$&$ 7.939\pm{1.149}$&$ 0.055\pm{0.008}$\\
UM 439                                      & 600 &$14200\pm{300  }$&$14000\pm{200  }$&$13700\pm{200  }$&$ 0.890\pm{0.047}$&$ 9.533\pm{0.579}$&$ 0.034\pm{0.006}$\\
TOL 0645-376                                & 350 &$12900\pm{1600}$&$13000\pm{1300}$&$12800\pm{1200}$&$ 3.272\pm{1.222}$&$ 7.342\pm{2.632}$&$ 0.254\pm{0.065}$\\
KISSR 1845, CG 903, HS 1440+4302            & 50  &$13300\pm{300  }$&$13300\pm{200  }$&$13100\pm{200  }$&$ 2.997\pm{0.286}$&$ 8.061\pm{0.622}$&$ 0.111\pm{0.011}$\\
MRK 5, 0635+756                             & 200 &$12200\pm{500  }$&$12400\pm{500  }$&$12200\pm{400  }$&$ 3.566\pm{0.515}$&$ 7.512\pm{0.951}$&$ 0.168\pm{0.016}$\\
II ZW 40                                    & 200 &$13300\pm{200  }$&$13300\pm{100  }$&$13100\pm{100  }$&$ 0.581\pm{0.026}$&$10.780\pm{0.523}$&$ 0.057\pm{0.003}$\\
MRK 1089, 0459-043                          & 100  &$11100\pm{700  }$&$11300\pm{700  }$&$11200\pm{700  }$&$ 5.533\pm{1.432}$&$ 6.302\pm{1.249}$&$ 0.519\pm{0.079}$\\

\enddata
\label{ionabun}
\end{deluxetable}

\clearpage
\begin{deluxetable}{lcccccc}
\tabletypesize{\scriptsize}
\setlength{\tabcolsep}{0.07in}
\renewcommand{\arraystretch}{1.5}
\rotate
\tablecolumns{7}
\tablewidth{0in}
\tablenum{6}
\tablecaption{ELEMENTAL ABUNDANCES}
\tablehead{
\colhead{Object Names} &
\colhead{Ref. for} &
\colhead{$12+\log(\frac{O}{H})$} & 
\colhead{$\log(\frac{N}{O})$} &
\colhead{$12+\log(\frac{O}{H})$} &
\colhead{$\log(\frac{N}{O})$} &
\colhead{Ref. for}\\
\colhead{\hfill} &
\colhead{$I(\lambda)$ \tablenotemark{\ }} &
\colhead{(present work)}&
\colhead{(present work)}&
\colhead{(literature)}&
\colhead{(literature)}&
\colhead{cols. 5 \& 6 \tablenotemark{\ }}}
\startdata
												
I ZW 18 NW, MRK 116, 0930+554                                  &4  &$7.181\pm{0.023}$&$-1.416\pm{0.049}$&$7.170\pm^{0.040}_{0.040}$&$-1.560\pm^{0.090  }_{0.090}$& 7\\
I ZW 18 SE, MRK 116, 0930+554                                  &4  &$7.280\pm{0.069}$&$-1.505\pm{0.070}$&$7.260\pm^{0.050}_{0.050}$&$-1.600\pm^{0.060  }_{0.060}$& 7\\
SBS 0335-052                                    &10 &$7.302\pm{0.005}$&$-1.408\pm{0.104}$&$7.290\pm^{0.010}_{0.010}$&$-1.580\pm^{0.030  }_{0.030}$&11\\
SBS 0940+544 N                                  &5  &$7.368\pm{0.006}$&$-1.382\pm{0.071}$&$7.315\pm^{0.016}_{0.016}$&$-1.523\pm^{0.068  }_{0.081}$& 7\\
SBS 1159+545                                    &5  &$7.415\pm{0.013}$&$-1.397\pm{0.055}$&$7.442\pm^{0.018}_{0.019}$&$-1.533\pm^{0.062  }_{0.072}$& 7\\
SBS 1415+437                                    &6  &$7.435\pm{0.008}$&$-1.423\pm{0.039}$&$7.504\pm^{0.014}_{0.014}$&$-1.568\pm^{0.042  }_{0.047}$& 7\\
KISSR 85                                        &12  &$7.488\pm{0.038}$&$-1.358\pm{0.078}$&$7.500\pm^{0.060}_{0.060}$&$-1.380\pm^{0.110  }_{0.110}$&12\\
UGC 4483, 0832+699                                    &5  &$7.495\pm{0.012}$&$-1.509\pm{0.041}$&$7.553\pm^{0.015}_{0.016}$&$-1.629\pm^{0.053  }_{0.060}$& 7\\
TOL 1214-277                                &1,3&$7.534\pm{0.023}$&$-1.324\pm{0.069}$&$7.576\pm^{0.025}_{0.026}$&$-1.449\pm^{0.080  }_{0.098}$& 7\\
KISSR 1490                                        &12  &$7.546\pm{0.027}$&$-1.300\pm{0.087}$&$7.560\pm^{0.070}_{0.070}$&$-1.410\pm^{0.130  }_{0.130}$&12\\
SBS 1211+540                                    &5  &$7.612\pm{0.012}$&$-1.458\pm{0.044}$&$7.671\pm^{0.010}_{0.011}$&$-1.601\pm^{0.058  }_{0.067}$& 7\\
KISSR 1013, KISSB 211                             &12  &$7.632\pm{0.033}$&$-1.095\pm{0.061}$&$7.660\pm^{0.050}_{0.050}$&$-1.180\pm^{0.090  }_{0.090}$&12\\
SBS 1249+493                                    &6  &$7.638\pm{0.034}$&$-1.468\pm{0.071}$&$7.703\pm^{0.024}_{0.025}$&$-1.584\pm^{0.082  }_{0.101}$& 7\\
116+583 B                                    &8  &$7.645\pm{0.048}$&$-1.326\pm{0.124}$&$7.680\pm^{0.050}_{0.050}$&$-1.450\pm^{0.110  }_{0.110}$&11\\
VII Zw 403, 1124+792                          &8  &$7.656\pm{0.013}$&$-1.394\pm{0.042}$&$7.690\pm^{0.010}_{0.010}$&$-1.530\pm^{0.030  }_{0.030}$&11\\
TOL 1304-353                                 &1  &$7.680\pm{0.034}$&$-1.316\pm{0.122}$&$7.721\pm^{0.033}_{0.036}$&$-1.442\pm^{0.127  }_{0.181}$& 7\\
SBS 1420+544, CG 413                                    &6  &$7.683\pm{0.016}$&$-1.429\pm{0.082}$&$7.728\pm^{0.014}_{0.014}$&$-1.542\pm^{0.078  }_{0.096}$& 7\\
SBS 1205+557                                     &8  &$7.705\pm{0.031}$&$-1.382\pm{0.056}$&$7.750\pm^{0.030}_{0.030}$&$-1.500\pm^{0.080  }_{0.080}$&11\\
SBS 1128+573                                    &8  &$7.721\pm{0.032}$&$-1.377\pm{0.090}$&$7.750\pm^{0.030}_{0.030}$&$-1.510\pm^{0.070  }_{0.070}$&11\\
C 1543+091                                  &1  &$7.721\pm{0.033}$&$-1.208\pm{0.117}$&$7.773\pm^{0.025}_{0.027}$&$-1.340\pm^{0.115  }_{0.156}$& 7\\
UM 461, SCHG 1148-020                                      &1  &$7.723\pm{0.024}$&$-1.417\pm{0.050}$&$7.764\pm^{0.019}_{0.020}$&$-1.579\pm^{0.059  }_{0.068}$& 7\\
I ZW 36, MRK 209, HARO 29,  1223+487         &8  &$7.735\pm{0.004}$&$-1.349\pm{0.039}$&$7.770\pm^{0.010}_{0.010}$&$-1.490\pm^{0.010  }_{0.010}$&11\\
SBS 1331+493 N                                  &5  &$7.747\pm{0.016}$&$-1.410\pm{0.048}$&$7.801\pm^{0.011}_{0.011}$&$-1.492\pm^{0.050  }_{0.056}$& 7\\
MRK 1434, SBS 1030+583                           &8  &$7.762\pm{0.009}$&$-1.479\pm{0.040}$&$7.790\pm^{0.010}_{0.010}$&$-1.600\pm^{0.020  }_{0.020}$&11\\
MRK 193, SBS 1152+579                                    &5  &$7.772\pm{0.011}$&$-1.310\pm{0.043}$&$7.865\pm^{0.010}_{0.010}$&$-1.274\pm^{0.047  }_{0.052}$& 7\\
MRK 71, NGC 2343, 0723+692 B                &8  &$7.787\pm{0.021}$&$-1.456\pm{0.049}$&$7.810\pm^{0.020}_{0.020}$&$-1.570\pm^{0.040  }_{0.040}$&11\\
MRK 600, 0248+042                           &10 &$7.791\pm{0.012}$&$-1.535\pm{0.040}$&$7.830\pm^{0.010}_{0.010}$&$-1.670\pm^{0.030  }_{0.030}$&11\\
MRK 71, NGC 2343, SBS 0749+568 B                                    &8  &$7.812\pm{0.049}$&$-1.332\pm{0.075}$&$7.850\pm^{0.050}_{0.050}$&$-1.440\pm^{0.110  }_{0.110}$&11\\
MRK 36, 1102+294                                      &1  &$7.815\pm{0.028}$&$-1.382\pm{0.111}$&$7.847\pm^{0.022}_{0.023}$&$-1.459\pm^{0.108  }_{0.144}$& 7\\
POX 4: C1148-203                            &1  &$7.817\pm{0.028}$&$-1.233\pm{0.052}$&$7.855\pm^{0.023}_{0.025}$&$-1.350\pm^{0.071  }_{0.085}$& 7\\
MRK 1416, SBC 0917+527                          &8  &$7.821\pm{0.018}$&$-1.507\pm{0.044}$&$7.860\pm^{0.020}_{0.020}$&$-1.620\pm^{0.040  }_{0.040}$&11\\
MRK 71, NGC 2343, 0723+692 A                &8  &$7.822\pm{0.004}$&$-1.373\pm{0.040}$&$7.850\pm^{0.010}_{0.010}$&$-1.520\pm^{0.010  }_{0.010}$&11\\
SBS 1533+574 A                                  &8  &$7.845\pm{0.034}$&$-1.303\pm{0.061}$&$7.880\pm^{0.030}_{0.030}$&$-1.430\pm^{0.080  }_{0.080}$&11\\
MRK 1486, SBS 1358+576                           &8  &$7.852\pm{0.015}$&$-1.179\pm{0.042}$&$7.880\pm^{0.010}_{0.010}$&$-1.310\pm^{0.030  }_{0.030}$&11\\
SBS 1331+493 A                                  &6  &$7.852\pm{0.058}$&$-1.408\pm{0.101}$&$7.893\pm^{0.047}_{0.053}$&$-1.495\pm^{0.089  }_{0.112}$& 7\\
KISSR 675, KISSB 187                              &12  &$7.861\pm{0.074}$&$-1.418\pm{0.140}$&$7.870\pm^{0.090}_{0.090}$&$-1.390\pm^{0.180  }_{0.180}$&12\\
FAIRALL 30                                   &1  &$7.875\pm{0.022}$&$-1.228\pm{0.046}$&$7.907\pm^{0.020}_{0.021}$&$-1.339\pm^{0.064  }_{0.076}$& 7\\
C 0840+120                                   &1  &$7.882\pm{0.028}$&$-1.347\pm{0.077}$&$7.915\pm^{0.025}_{0.027}$&$-1.448\pm^{0.078  }_{0.095}$& 7\\
SBS 0926+606                                     &8  &$7.884\pm{0.015}$&$-1.349\pm{0.042}$&$7.910\pm^{0.010}_{0.010}$&$-1.470\pm^{0.030  }_{0.030}$&11\\
I ZW 49, MRK 59, NGC 4861, ARP 209, 1256+351                                    &13  &$7.889\pm{0.046}$&$-1.290\pm{0.075}$&$7.940\pm^{0.030}_{0.030}$&$-1.440\pm^{0.050  }_{0.060}$& 7\\
KISSR 1194                                        &12  &$7.898\pm{0.028}$&$-1.367\pm{0.068}$&$7.920\pm^{0.040}_{0.040}$&$-1.460\pm^{0.080  }_{0.080}$&12\\
KISSR 396, KISSB 145, WAS 81                      &12  &$7.906\pm{0.030}$&$-1.508\pm{0.058}$&$7.920\pm^{0.040}_{0.040}$&$-1.580\pm^{0.070  }_{0.070}$&12\\
UM 420, 0218+003                            &10 &$7.911\pm{0.054}$&$-0.979\pm{0.087}$&$7.930\pm^{0.050}_{0.050}$&$-1.080\pm^{0.120  }_{0.120}$&11\\
1437+370                                    &5  &$7.914\pm{0.021}$&$-1.400\pm{0.050}$&$7.965\pm^{0.014}_{0.014}$&$-1.488\pm^{0.053  }_{0.060}$& 7\\
KISSR 1778                                        &12  &$7.915\pm{0.067}$&$-1.351\pm{0.131}$&$7.930\pm^{0.080}_{0.080}$&$-1.440\pm^{0.160  }_{0.160}$&12\\
UM 462, SCHG 1150-021                                     &1  &$7.923\pm{0.015}$&$-1.412\pm{0.071}$&$7.960\pm^{0.017}_{0.018}$&$-1.506\pm^{0.080  }_{0.098}$& 7\\
SBS 1222+614                                    &8  &$7.935\pm{0.009}$&$-1.508\pm{0.039}$&$7.950\pm^{0.010}_{0.010}$&$-1.610\pm^{0.020  }_{0.020}$&11\\
TOL 1304-386                                &1  &$7.947\pm{0.025}$&$-1.082\pm{0.055}$&$7.982\pm^{0.020}_{0.021}$&$-1.181\pm^{0.070  }_{0.084}$& 7\\
1533+469                                    &6  &$7.953\pm{0.038}$&$-1.183\pm{0.068}$&$8.042\pm^{0.027}_{0.029}$&$-1.354\pm^{0.064  }_{0.075}$& 7\\
SBS 0907+543                                    &8  &$7.957\pm{0.033}$&$-1.461\pm{0.083}$&$8.010\pm^{0.030}_{0.030}$&$-1.590\pm^{0.080  }_{0.080}$&11\\
UM 469                                      &1,3&$7.960\pm{0.095}$&$-1.052\pm{0.156}$&$7.982\pm^{0.068}_{0.080}$&$-1.113\pm^{0.125  }_{0.177}$& 7\\
1054+365, CG 798                            &8  &$7.965\pm{0.015}$&$-1.380\pm{0.044}$&$7.970\pm^{0.020}_{0.020}$&$-1.480\pm^{0.030  }_{0.030}$&11\\
MRK 1450, SBS 1135+581                                    &5  &$7.965\pm{0.013}$&$-1.363\pm{0.041}$&$8.007\pm^{0.013}_{0.013}$&$-1.412\pm^{0.051  }_{0.058}$& 7\\
MRK 1271, TOL 1053+064                          &10 &$7.975\pm{0.013}$&$-1.257\pm{0.042}$&$7.990\pm^{0.010}_{0.010}$&$-1.380\pm^{0.030  }_{0.030}$&11\\
NGC 5253-6                                   &2  &$7.981\pm{0.047}$&$-1.136\pm{0.089}$&$8.006\pm^{0.041}_{0.045}$&$-1.212\pm^{0.078  }_{0.096}$& 7\\
MRK 1304, UM 448, SCHG 1139+006                            &10 &$7.984\pm{0.048}$&$-0.929\pm{0.092}$&$7.990\pm^{0.040}_{0.040}$&$-1.010\pm^{0.100  }_{0.010}$&11\\
MRK 22, SBS 0946+558                                    &5  &$7.984\pm{0.016}$&$-1.400\pm{0.044}$&$8.026\pm^{0.015}_{0.015}$&$-1.495\pm^{0.052  }_{0.059}$& 7\\
1441+294, CG 1258                           &8  &$7.986\pm{0.061}$&$-1.318\pm{0.103}$&$7.990\pm^{0.060}_{0.060}$&$-1.410\pm^{0.130  }_{0.130}$&11\\
KISS 49, KISSB 94, CG 177                       &12  &$7.988\pm{0.060}$&$-1.240\pm{0.115}$&$8.000\pm^{0.060}_{0.060}$&$-1.300\pm^{0.110  }_{0.110}$&12\\
MRK 1409, SBS 0741+535                         &8  &$7.991\pm{0.044}$&$-1.448\pm{0.078}$&$8.010\pm^{0.040}_{0.040}$&$-1.540\pm^{0.100  }_{0.100}$&11\\
0948+532                                    &5  &$7.995\pm{0.022}$&$-1.339\pm{0.050}$&$8.032\pm^{0.015}_{0.015}$&$-1.442\pm^{0.053  }_{0.061}$& 7\\
TOL 1345-420                                &1  &$7.999\pm{0.051}$&$-1.530\pm{0.092}$&$8.033\pm^{0.038}_{0.041}$&$-1.618\pm^{0.091  }_{0.116}$& 7\\
UM 439                                      &1  &$8.018\pm{0.024}$&$-1.385\pm{0.085}$&$8.045\pm^{0.020}_{0.020}$&$-1.444\pm^{0.090  }_{0.114}$& 7\\
TOL 0645-376                                &1  &$8.026\pm{0.119}$&$-1.076\pm{0.200}$&$8.046\pm^{0.083}_{0.102}$&$-1.133\pm^{0.148  }_{0.226}$& 7\\
KISSR 1845, CG 903, HS 1440+4302                  &12  &$8.044\pm{0.027}$&$-1.397\pm{0.071}$&$8.050\pm^{0.040}_{0.040}$&$-1.470\pm^{0.080  }_{0.080}$&12\\
MRK 5, 0635+756                             &10 &$8.044\pm{0.042}$&$-1.292\pm{0.083}$&$8.040\pm^{0.040}_{0.040}$&$-1.360\pm^{0.100  }_{0.100}$&11\\
II ZW 40                                    &1  &$8.055\pm{0.020}$&$-0.974\pm{0.048}$&$8.046\pm^{0.025}_{0.026}$&$-1.042\pm^{0.095  }_{0.121}$& 7\\
MRK 1089, 0459-043                          &10 &$8.073\pm{0.070}$&$-0.993\pm{0.135}$&$8.040\pm^{0.060}_{0.060}$&$-1.050\pm^{0.140  }_{0.140}$&11\\
\enddata
\tablenotetext{\ }{REFERENCES.--(1)\citealt{campbell86};(2)\citealt{walsh89};(3)\citealt{pagel92};(4)\citealt{skillman93};(5)\citealt{izotov94};(6)\citealt{thuan95};(7)\citealt{kobulnicky96};(8)\citealt{izotov97}; (9)\citealt{izotov98a}; (10)\citealt{izotov98b}; (11)\citealt{izotov99};(12)\citealt{melbourne04};(13)\citealt{kobulnicky98}.}\label{bigtable}
\end{deluxetable}
\clearpage
\clearpage
\clearpage
\clearpage
\clearpage
\clearpage
\clearpage
\clearpage
\begin{figure}
\epsscale{.80}
\plotone{f8.eps}
\caption{Difference log[ \niio~] - log[ $\left<N^+/O^+\right>$ ] as a function of log[ $\left<N^+/O^+\right>$ ] for our \cloudy~models, where log[ \niio~] is based on line flux ratios while log[ $\left<N^+/O^+\right>$ ] is based on eq. [\ref{eq:meanifrac}].\label{icf}}
\end{figure}
\clearpage
\begin{figure}
\epsscale{.80}
\plotone{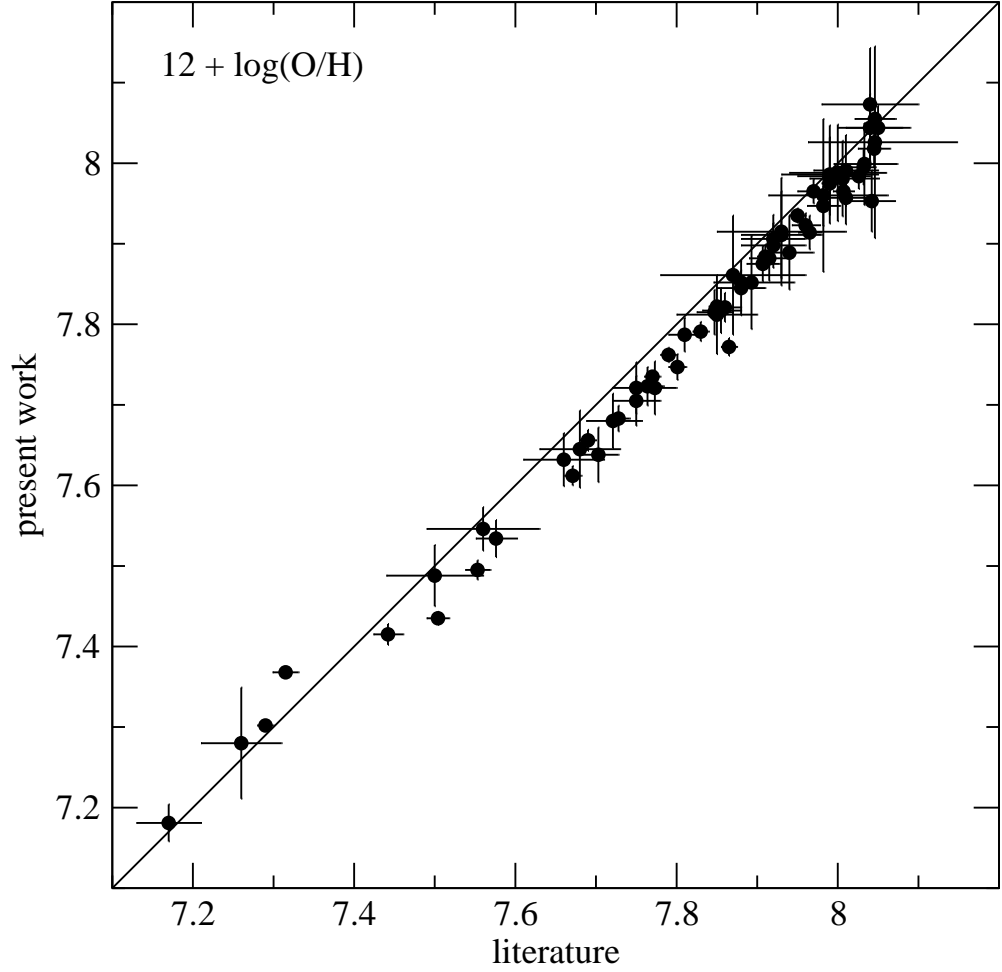}
\caption{Present work vs. published \oiih~values for our sample of 68 low metallicity systems. The literature values were taken from the sources indicated in the footnote of Table~\ref{bigtable}. The diagonal represents points of one to one correspondance.\label{oldnew_o2h}}
\end{figure}
\clearpage
\begin{figure}
\epsscale{.80}
\plotone{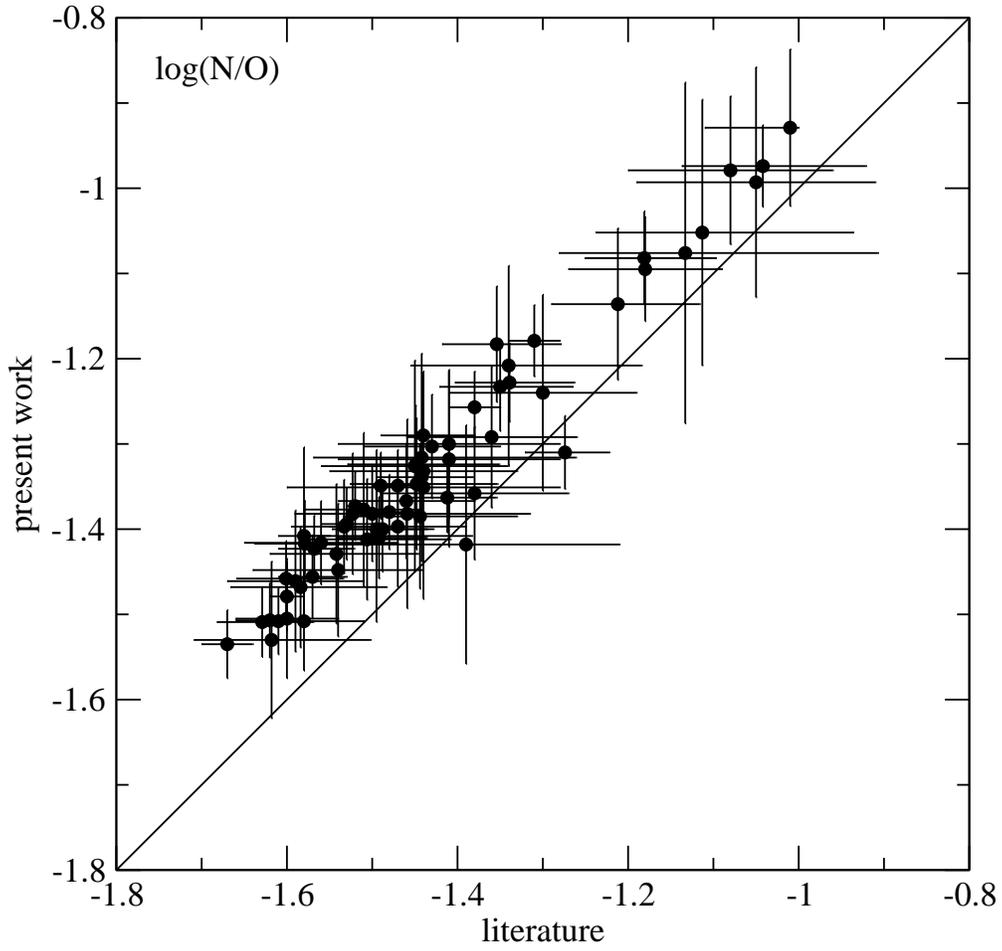}
\caption{Same as Fig.~\ref{oldnew_o2h} but for log(N/O).\label{oldnew_n2o}}
\end{figure}
\clearpage
\begin{figure}
\epsscale{.80}
\plotone{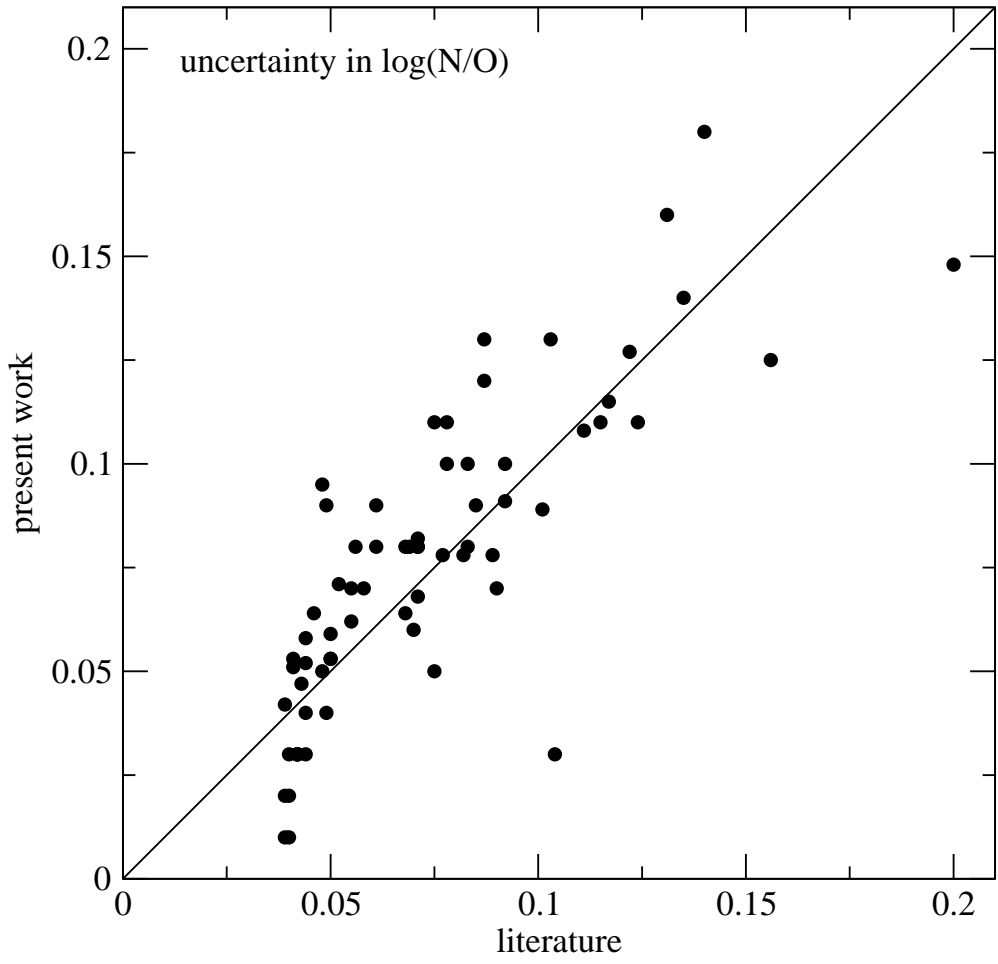}
\caption{Same as Fig.~\ref{oldnew_o2h} but for the uncertainty in log(N/O).\label{oldnew_sigma_n2o}}
\end{figure}
\clearpage
\begin{figure}
\epsscale{.80}
\rotatebox{ 270 }{ \plotone{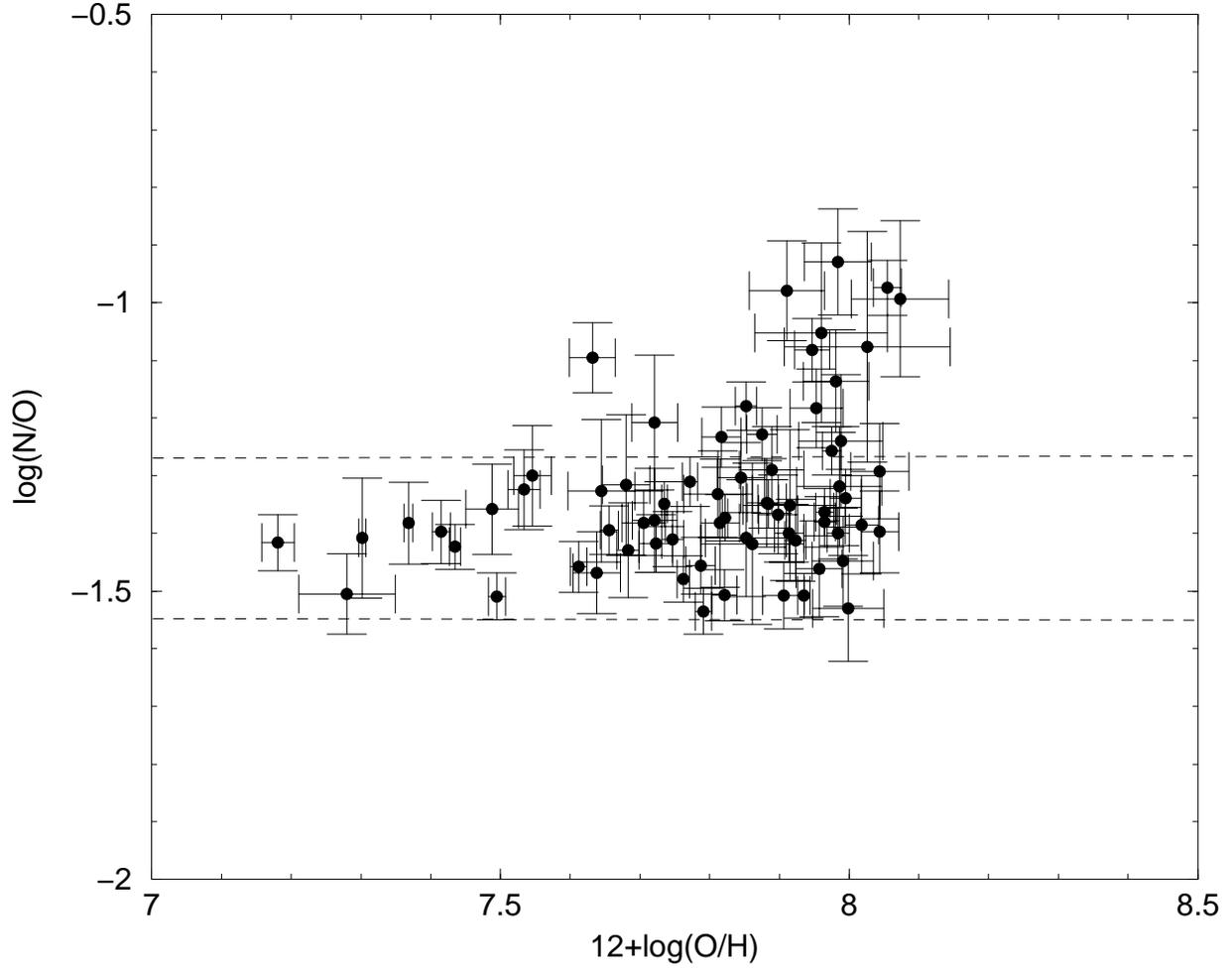} }
\caption{Log(N/O) versus \oiih~ as determined in this study for our sample objects. The dashed lines show the upper and lower limits for the N/O plateau as we define them (see the discussion of Fig.~\ref{n2ohisto}).\label{oursample}}
\end{figure}
\clearpage
\begin{figure}
\epsscale{.80}
\rotatebox{ -90 }{ \plotone{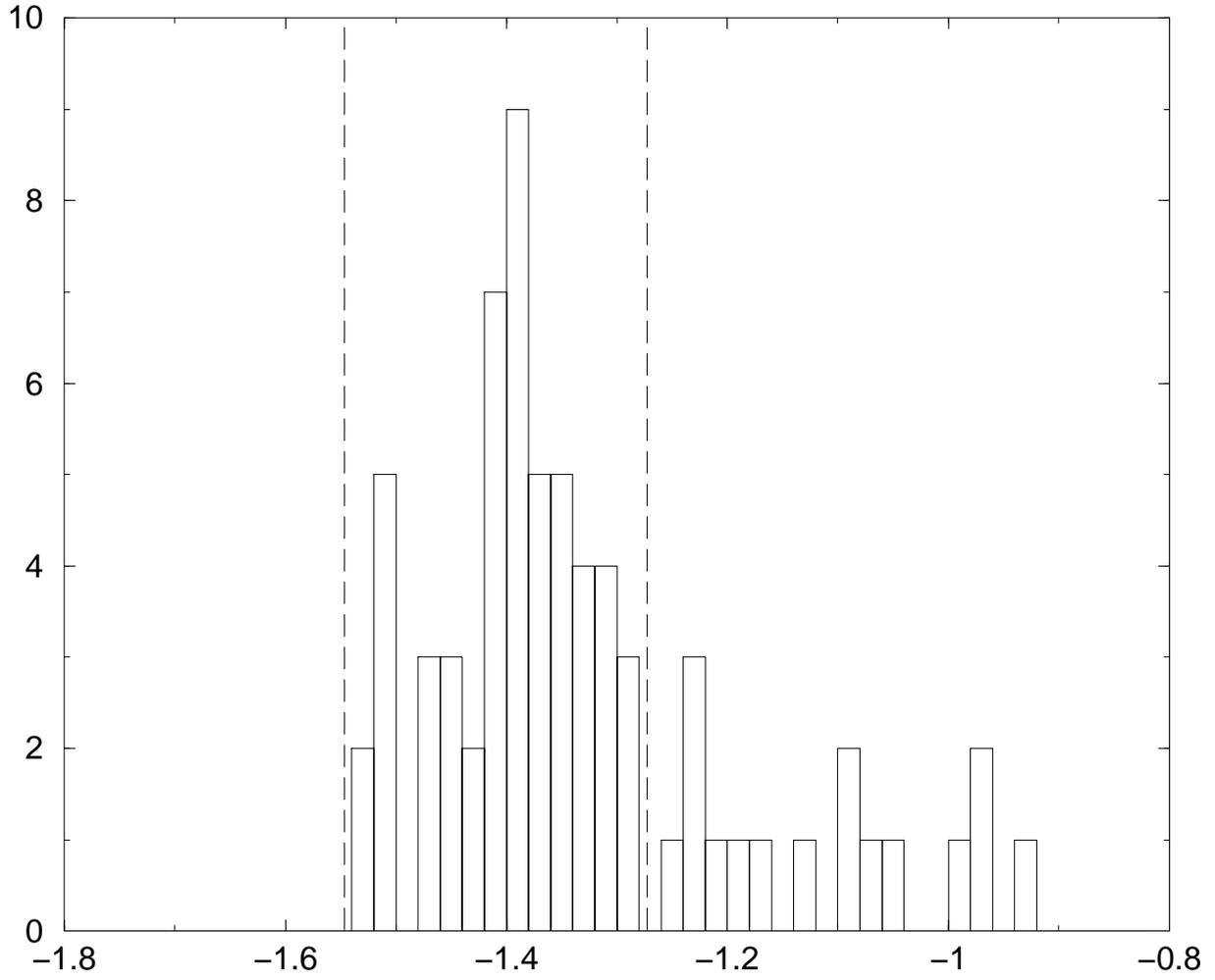} }
\caption{Observed distribution of values of log(N/O) as determined in this paper. The vertical dashed lines show the upper and lower limits for the N/O plateau.\label{n2ohisto}}
\end{figure}

\end{document}